\documentclass[nopreprintline]{elsarticle}

\usepackage{graphicx}
\usepackage{mathrsfs}
\usepackage{array,multirow,graphicx}
\usepackage{siunitx}
\usepackage[colorlinks=true, allcolors=blue]{hyperref}
\usepackage{mathtools}
\usepackage{booktabs}
\usepackage{amsmath,amsthm,amsfonts,amssymb}
\usepackage[labelfont=bf]{caption}
\usepackage{tabularx}
\usepackage{algorithm}
\usepackage[noend]{algpseudocode}
\MakeRobust{\Call}

\usepackage{lineno}

\DeclareMathOperator*{\argmin}{arg\,min}
\newcommand{\norm}[1]{\left\lVert#1\right\rVert}

\date{}

\begin{document}
\begin{frontmatter}

\author[1]{Cody Rucker\corref{c1}}
\cortext[c1]{Corresponding author}
\ead{crucker@uoregon.edu}

\author[1,2]{Brittany A. Erickson}

\affiliation[1]{organization={Department of Computer Science, University of Oregon},
addressline={1477 E 13th Ave},
postcode={97403},
city={Eugene},
country={United States of America}}

\affiliation[2]{organization={Department of Earth Sciences, University of Oregon},
addressline={1272 University of Oregon},
city={Eugene},
postcode={97403},
country={United States of America}}

\title{Physics-Informed Deep Learning of Rate-and-State Fault Friction}

\begin{abstract}
Direct observations of earthquake nucleation and propagation are few and yet the next decade will likely see an unprecedented increase in indirect, surface observations that must be integrated into modeling efforts. Machine learning (ML) excels in the presence of large data and is an actively growing field in seismology. However, not all ML methods incorporate rigorous physics, and purely data-driven models can predict physically unrealistic outcomes due to observational bias or extrapolation. Our work focuses on the recently emergent Physics-Informed Neural Network (PINN), which seamlessly integrates data while ensuring that model outcomes satisfy rigorous physical constraints. In this work we develop a multi-network PINN for both the forward problem as well as for direct inversion of nonlinear fault friction parameters, constrained by the physics of motion in the solid Earth, which have direct implications for assessing seismic hazard. We present the computational PINN framework for strike-slip faults in 1D and 2D subject to rate-and-state friction. Initial and boundary conditions define the data on which the PINN is trained. While the PINN is capable of approximating the solution to the governing equations to low-errors, our primary interest lies in the network's capacity to infer friction parameters during the training loop. We find that the network for the parameter inversion at the fault performs much better than the network for material displacements to which it is coupled. Additional training iterations and model tuning resolves this discrepancy, enabling a robust surrogate model for solving both forward and inverse problems relevant to seismic faulting. 
\end{abstract}


\end{frontmatter}




\section{Context and Motivation}\label{sec: intro}

Faults are home to a vast spectrum of event types, from slow aseismic creep, to slow-slip to megathrust earthquakes followed by postseismic afterslip. The Cascadia subduction zone in the Pacific Northwest, for example, hosts several types of slow earthquake processes including low (and very low) frequency earthquakes, non-volcanic tremor  (NVT) and slow-slip events (SSE) \citep{Ide2007}, but also large, fast earthquakes, the last of which was a magnitude $\sim$9 in the year 1700 \citep{Atwater2005}.  Understanding the physical mechanisms for such diversity of slip styles is crucial for mitigating the associated hazards but major uncertainties remain in the depth-dependency of frictional properties at fault zones, which affect fault locking and therefore rupture potential \citep{Brodsky2020, NAP25761}. Direct observations of earthquake nucleation and propagation are few and yet the next decade will likely see an unprecedented increase in indirect, surface observations that could be integrated into modeling efforts \citep{Bergen2019}.  
 
 Traditional numerical approaches for solving the partial differential equations (PDE) governing earthquake processes (e.g. finite difference methods) have seen incredible growth in the past century, in particular in terms of convergence theory and high-performance computing. Traditional methods employ a mesh (either a finite number of grid points/nodes or elements) and a range of time-integration schemes in order to obtain an approximate solution whose accuracy depends directly on the mesh size (with error decreasing with decreasing node spacing or element size). This mesh dependency introduces limitations when high resolution is needed, and while traditional methods give rise to a forward problem, solving inverse problems require additional machinery and can be prohibitively expensive \citep{kern2016numerical, givoli2021tutorial}.  In addition, noisy or sparse data cannot be seamlessly integrated into the computational framework of traditional methods.

Machine learning (ML), on the other hand, excels in the presence of large data and is an actively growing field in seismology, with applications ranging from earthquake early warning (EEW) to ground‐motion prediction \citep{Kong2018}. However,  not all ML methods incorporate rigorous physics, and purely data-driven models can predict physically unrealistic outcomes due to observational bias or extrapolation \citep{Zhao2023}. A new Deep Learning technique has recently emerged called the Physics-Informed Neural Network (PINN), which seamlessly integrates sparse and/or noisy data while ensuring that model outcomes satisfy rigorous physical constraints. PINNs do not outperform traditional numerical methods for forward problems (except in high-dimensional settings) \citep{karniadakis2021physics}, but they offer advantages over traditional numerical methods in that both forward and inverse problems can be solved in the same computational framework. However, the majority of PINN applications are currently limited to simple mechanical models, forward problems and/or do not incorporate real-world observations \citep{Fukushima2023, Okazaki2022, KARIMPOULI20201993}. Here we introduce a new, physically-rigorous modeling framework for both forward and inverse problems that can be integrated with observational data in order to better understand earthquake fault processes.  This Deep Learning approach will vastly expand our computational abilities to explore and infer relevant parameter spaces responsible for slip complexity, and the conditions that enable the world's largest earthquakes. 

 Though the PINN framework lacks the robust error analysis that comes with traditional methods, a large number of publications have emerged since $\sim$2017 which aim to customize PINNs through the use of different activation functions, gradient optimization techniques, neural network architecture, and loss function structure \cite{cuomo2022scientific}. Careful formulation of the loss function using the weak form of the PDE have been proposed for constructing deep learning analogues of Ritz \citep{yu2018deep} and Galerkin \citep{kharazmi2019variational, kharazmi2021hp, jagtap2020conservative} methods which use numerical quadrature to reduce the order of the PDE resulting in a simpler learning problem \citep{ciarlet2002finite, ern2004theory}. In tandem, statistical learning theory has been used to deduce global error bounds for PINNs in terms of optimization error, generalization error, and approximation error \cite{kutyniok2022mathematics}. For wide but shallow networks utilizing hyperbolic tangent activation functions, the approximation error has been shown to be bounded over Sobolev spaces \cite{de2021approximation}. Bounds on PINN generalization error have been derived for linear second-order PDE \cite{shin2020convergence} (later extended to all linear problems \cite{shin2020error}) and some specific cases like Navier-Stokes \cite{de2022error}. Moreover, the abstract framework for PINNs can leverage stability of a PDE to provide conditions under which generalization error is small whenever training error is small for both forward and inverse problems \cite{mishra2022estimates_forward, mishra2022estimates_inverse}. More recently, a PINN-specific optimization algorithm has achieved markedly improved accuracy over other optimization algorithms by incorporating a PDE energy into the backpropagation step \cite{muller2023achieving}. In addition to this rapid framework development, PINNs have been shown to perform well on a variety of physical problems like Navier-Stokes \cite{wang2020multi, sun2020surrogate, jin2021nsfnets}, convection heat transfer \cite{cai2021physics}, solid mechanics, \cite{haghighat2021physics, goswami2020transfer} and the Euler equations \cite{mao2020physics}.

In this work we focus specifically on rate-and-state friction \cite[e.g.][]{scholz_2019}, an experimentally-motivated, nonlinear friction law capable of reproducing a wide range of observed earthquake behaviors and is used in nearly all modern dynamic rupture and earthquake cycle simulations \citep{Harris2009, Erickson2020}. A better understanding of the depth-dependency of rate-and-state parameters - which have a direct correlation to fault locking and seismic rupture potential -- is a fundamental task \citep{Brodsky2020, NAP25761}. To address this task we develop a multi-network PINN for modeling a vertical, strike-slip frictional fault embedded in an elastic half-space, and consider deformation in both 1D and 2D. The paper is organized as follows: In section \ref{sec: PINN framework} we first provide an overview of the physics-informed deep learning framework and PINN architecture for general initial-boundary-value problems, in order to best describe the implementation to our application problem. In section \ref{sec: governing eqns} we provide specific details of the PINN framework applied problems, first illustrated in 1D with an example forward problem, then further developed to include inverse problems in 2D. In section \ref{sec: verification} we report details of our optimal network architecture and training methods, verifying our methods with a manufactured solution to ensure accuracy of our inversions. We conclude with a summary and discussion of future work in section \ref{sec: summary}.

\section{Physics-Informed Deep Learning Framework}\label{sec: PINN framework}
The physics governing motion in many applications in science and engineering give rise to partial differential equations (PDE) where analytic solutions can be difficult to obtain (due, e.g. to complex material properties, boundary conditions, geometry) and we commonly turn to numerical methods. The Physics-Informed Neural Network (PINN) is a deep learning (DL) framework for approximating solutions to PDEs. Though this DL framework lacks the robust mathematical theory of the traditional methods, it shows particular promise in solving problems for which traditional numerical methods are ill-suited\citep{raissi2019physics}. The DL framework produces a closed, analytic form for the solution, which is continuous and defined at every point in the domain allowing one to evaluate the solution ``off-grid" without having to resolve the PDE (i.e. it is mesh-free) \citep{raissi2019physics}. In addition, both forward and inverse problems can be solved in the same computational setting, as will be shown in subsequent sections.

\subsection{Feed-forward Deep Neural Networks}

PINNs are extensions of a general feed-forward neural network. We let $\mathbf{x}\in \mathbb{R}^n$ and define a weighting matrix $W\in \mathbb{R}^{m\times n}$ and bias vector $b\in \mathbb{R}^m$. A single hidden layer of a neural network can be expressed as 
\begin{align}
    \ell(\mathbf{x}; \theta) = \varphi(W\mathbf{x} + b), \quad \text{where }\theta = (W, b),
\end{align}
and $\varphi$ is a known (nonlinear) activation function. Deep neural networks are obtained by repeated composition of hidden layers \citep{kollmannsberger2021deep}.

We let positive integer $L$ be the deep neural network depth (i.e. number of hidden layers) and let $\{\varphi_i\}_{i=1}^L$ be a collection of activation functions along with a sequence of trainable network parameters $\{\theta_i\}_{i=0}^{L}$ where $\theta_k=(W_k, b_k)$ for each $0\leq k\leq L$. Here we assume each layer consists of $m$ neurons but this simplifying assumption may be omitted so long as one ensures that each weight and bias are of the correct dimension for matrix multiplication and addition, respectively. If $\mathbf{x}\in \mathbb{R}^n$ is the network input and network parameters $(W_0, b_0)\in \mathbb{R}^{m\times n}\times \mathbb{R}^m$, $(W_k, b_k)\in \mathbb{R}^{m\times m}\times \mathbb{R}^m$ for $1 \leq k<L$, and $(W_L, b_L)\in \mathbb{R}^{d\times m}\times \mathbb{R}^d$, then the recursive composition 
\begin{subequations}
    \begin{align}
	   \ell_0 &= \mathbf{x},\\
	   \ell_k &= \varphi_k (W_k \ell_{k-1} + b_k) \quad \text{ for } 0<k< L,
    \end{align}
    \label{Defn: recursive_NNet_def}%
\end{subequations}
defines a feed-forward, deep neural network $\mathcal{N}:\mathbb{R}^n \to \mathbb{R}^d$ (of depth $L$ and width $m$) with network parameters $\theta$ by $\mathcal{N}(\mathbf{x};\theta) = W_L\ell_{L-1} + b_L$. 

\subsection{PINN Architecture}

The PINN architecture is defined by extending the feed-forward deep neural network to enforce physical conditions set by an initial-boundary value problem (IBVP).  For simplicity in the notation (as well as generality) we define this extension for a general IBVP, with specific details for our application given in the subsequent section.  Consider the general operator form of an IBVP given by 
\begin{subequations}
    \begin{alignat}{2}
        \mathcal{L}\left[{\bf u}; \lambda\right] &= {\bf k}, \quad && \text{ in } \widehat{\Omega} , \label{eqn: ibvp(operator form)a}\\
        \mathcal{B}\left[{\bf u}; \lambda\right] &= {\bf g}, \quad &&\text{ on } \partial \widehat{\Omega},\label{eqn: ibvp(operator form)b}
    \end{alignat}
    \label{eqn: ibvp(operator form)}%
\end{subequations}
where $\widehat{\Omega}\subseteq \mathbb{R}^n$, with boundary (including internal interfaces, like faults) $\partial \widehat{\Omega}$. Vector ${\bf u}$ is the unknown solution and ${\bf g}$ is boundary data.  The source term ${\bf k}$ encompasses external and internal body forces, and $\mathcal{L}$, $\mathcal{B}$ are differential and boundary operators parameterized by $\lambda$.

%

\begin{figure}[H]
    \centering
    \includegraphics[width=\textwidth]{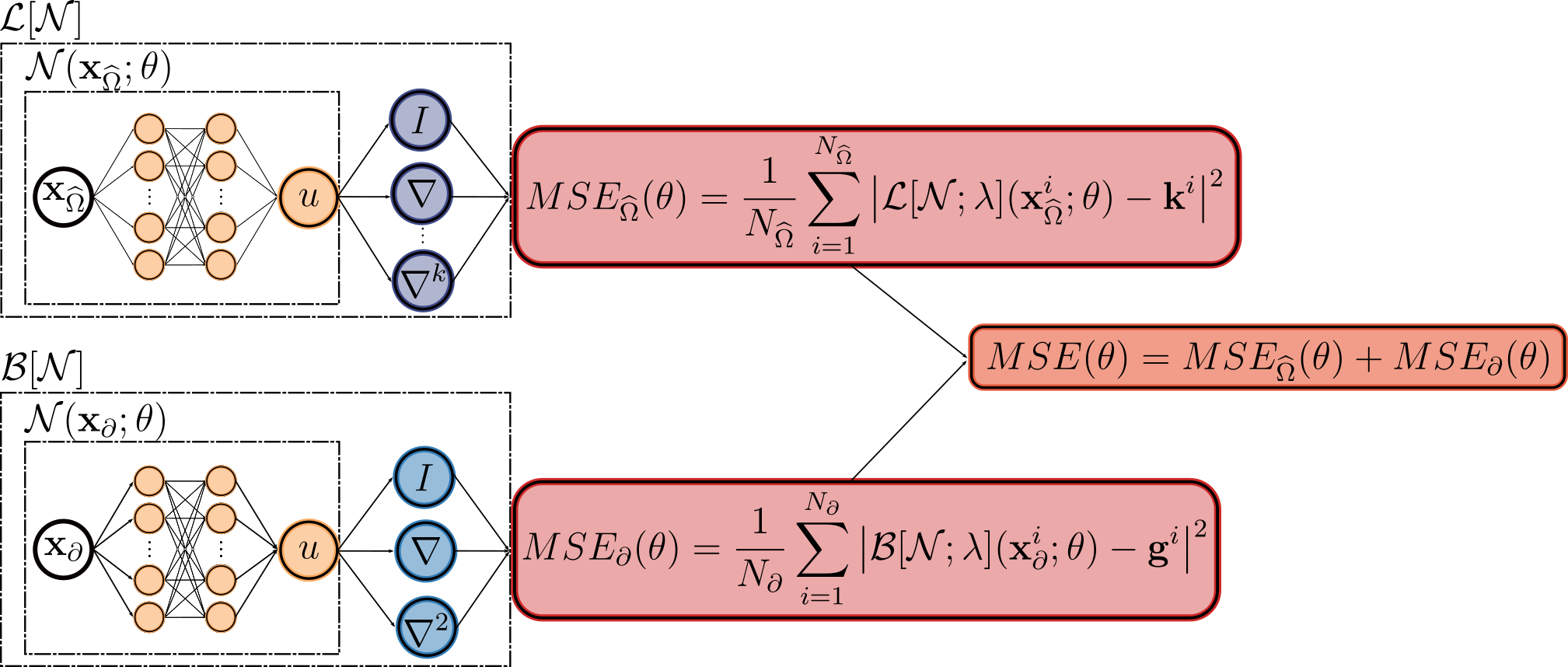}
    \caption{A schematic of the PINN framework for solving the general boundary value problem from equation~\eqref{eqn: ibvp(operator form)}. Displacement approximation network $\mathcal{N}$ is trained on interior and boundary subdomains which are governed by operators $\mathcal{L}$ and $\mathcal{B}$, respectively.}
    \label{fig:PINN diagram}%
\end{figure}

The PINN first approximates the solution to the IBVP \eqref{eqn: ibvp(operator form)} using a feed-forward, deep neural network, i.e. we assume ${\bf u}({\bf x})\approx \mathcal{N}({\bf x}; \theta)$ and consider $N_\partial$ many boundary points $\{{\bf x}^i_\partial, {\bf g}^i\}_{i=1}^{N_\partial}$ where ${\bf g}^i={\bf g}({\bf x}^i_\partial)$ for each $1\leq i \leq N_\partial$, and a set of $N_{{\widehat{\Omega}}}$ internal collocation points $\{ {\bf x}_{{\widehat{\Omega}}}^i, {\bf k}^i\}_{i=1}^{N_{\widehat{\Omega}}}$ with ${\bf k}^i={\bf k}({\bf x}^i_{\widehat{\Omega}})$ for $1\leq i \leq N_{\widehat{\Omega}}$. Throughout this work, collocation points are generated from uniform random sampling. From \eqref{eqn: ibvp(operator form)} we construct an objective function $MSE$ (based on the mean-square error) given by
\begin{align}
    MSE(\theta) = MSE_{{\widehat{\Omega}}}(\theta)+  MSE_\partial(\theta) ,
    \label{eqn: MSE}%
\end{align}
where
\begin{subequations}
    \begin{align}
        MSE_{{\widehat{\Omega}}}(\theta) &= \frac{1}{N_{{\widehat{\Omega}}}}\sum_{i=1}^{N_{{\widehat{\Omega}}}} \big |\mathcal{L}[\mathcal{N};\lambda]({\bf x}^i_{{\widehat{\Omega}}}; \theta) - {\bf k}^i \big |^2,\\[-5pt]
        MSE_\partial(\theta) &= \frac{1}{N_\partial}\sum_{i=1}^{N_\partial} \big |\mathcal{B}[\mathcal{N}; \lambda]({\bf x}^i_\partial; \theta) - {\bf g}^i \big |^2,
    \end{align}
    \label{eqn: MSE loss}%
\end{subequations}
are the contributions to the total loss from the PDE \eqref{eqn: ibvp(operator form)a} and boundary conditions \eqref{eqn: ibvp(operator form)b}, respectively, corresponding to misfit of the network with data. Because our objective function \eqref{eqn: MSE} consists of multiple competing loss functions it is referred to as a multi-objective loss. Solving \eqref{eqn: ibvp(operator form)} is done by minimizing \eqref{eqn: MSE} with respect to the network parameters $\theta$, typically done via an optimization algorithm that approximates the set of solution parameters $\theta^*$ such that $\theta^* = \argmin_\theta MSE(\theta)$.  Figure~\ref{fig:PINN diagram} gives a schematic representation of the objective function \eqref{eqn: MSE}. Note that for the rest of this work with omit the notation specifying MSE dependency on network parameters $\theta$ and assume this is understood. 
%


\section{Learning Problems for Earthquakes on Rate-and-State Faults}\label{sec: governing eqns}

In order to best illustrate the PINN computational set-up, we begin by considering a relevant problem in 1D, in order to introduce our methodology in a simplified framework.  Here we focus on the forward problem, with details of the inverse problem included in the subsequent section on the 2D application problem.

\subsection{1D Illustration}
The solid Earth is governed by momentum balance and a constitutive relation defining the material rheology; in this work we assume elastic material properties. These assumptions give rise to the elastic wave equation, namely 
\begin{equation}
 u_{tt} = c^2 u_{xx} + s(x, t), \quad x \in [0, 1], t \geq 0,
    \label{eqn: 1Delastodynamic}%
\end{equation}
where $u(x, t)$ is the Earth's material displacement, $c = \sqrt{\mu/\rho}$ is the wavespeed where $\rho$ and $\mu$ are the the density and shear modulus, and $s(x,t)$ accounts for external and/or body forces.  Assumed initial and boundary conditions are given by 
\begin{subequations}
    \begin{alignat}{2}
    u(x, 0) &= u_0(x), \\
    u_t(x, 0) &= v_0(x)\\
        -\mu u_x &= F + g_0(t), \quad &&x=0, \label{eqn: 1D fault condition}\\
        Zu_t + \mu u_x &= g_1(t), \quad &&x=1,\label{eqn: 1D fault condition2}
    \end{alignat}
    \label{eqn: 1D boundary conditions}%
\end{subequations}
where $Z = \sqrt{\mu\rho}$ is the shear impedance. If boundary data $g_0(t) = g_1(t) = 0$, equation \eqref{eqn: 1D fault condition} corresponds to the requirement that fault shear stress be equal to frictional strength $F$, and \eqref{eqn: 1D fault condition2} allows waves to freely exit the domain. 

In this work we consider rate-and-state dependent friction (RSF), an experimentally-motivated, nonlinear friction law, used in the majority of modern earthquake simulations for its ability to reproduce a wide range of observed seismic and aseismic behaviors \cite{dieterich1979, Ruina1983, Marone1998}. 
In this context, the frictional strength is given by 
\begin{equation}
       F = \bar{\sigma}_n f(V, \psi),  \label{eqn: strength} 
\end{equation}
where $V$ is the slip rate (jump in velocity across a fault interface), $\bar{\sigma}_n$ is the effective normal stress (normal stress minus pore fluid pressure), and $f$ is a friction coefficient given by 
\begin{equation}\label{eqn: 1D RNS fritction}
       f(V, \psi) = a\ln\left(\frac{V}{V_0} \right) + \psi,
\end{equation}
where $V_0$ is a reference slip rate  and material parameter $a$ is the ``direct effect" \citep{VANDENENDE2018273}. State variable $\psi$ evolves from some initial value $\psi_0$ according to its own evolution law
\begin{subequations}
\begin{align}
    \frac{d \psi}{dt} &= G(V, \psi) + h(t), \label{eqn: aging}\\
    \psi(0) &= \psi_0
\end{align}
\end{subequations}
where we assume the aging law, i.e. $G(V, \psi) = (bV_0/D_c)\exp{(\frac{f_0-\psi}{b} - \frac{|V|}{V_0}})$, which allows state to evolve even in the absence of slip \cite{Marone1998}, and $h(t)$ allows for the incorporation of additional source data. Here $D_c$ is the characteristic slip distance, $f_0$ is a reference friction coefficient for sliding at speed $V_0$ and material parameter $b$ captures time-dependent ``evolution" effects \citep{VANDENENDE2018273}.

%
%
%

The initial boundary value problem (IBVP) formed by the PDE \eqref{eqn: 1Delastodynamic}, specified initial/boundary conditions \eqref{eqn: 1D boundary conditions} and state-evolution equation \eqref{eqn: aging} correspond to specific differential and boundary operators $\mathcal{L}$ and $\mathcal{B}$ defined in 
\eqref{eqn: ibvp(operator form)}, for example, here $\mathcal{L} =\partial^2 / \partial t^2 - c^2 \partial^2 / \partial x^2$. Assuming all parameters of the IBVP are known, one can solve the forward problem for the unknown material displacement $u(x, t)$ and state variable $\psi(t)$. We define neural networks $\mathcal{N}(x, t; \theta) \approx u(x, t)$ and $\mathcal{N}^\psi(t; \theta_\psi) \approx \psi(t)$ to approximate the exact solutions. The PDE, boundary and initials conditions all give rise to a term in the objective function. We refer to these terms as the component loss functions and define them as  
\begin{subequations}
    \begin{align}
        MSE_\Omega &= \frac{1}{N_\Omega}\sum_{i=1}^{N_\Omega} \big |\mathcal{N}_{tt} - c^2\mathcal{N}_{xx}  -s\big |^2, \\
        MSE_0 &= \frac{1}{N_0} \sum_{i=1}^{N_0} \big | -\mu \mathcal{N}_x - F - g_0 \big |^2, \label{eqn: fault loss1D}\\
        MSE_1 &= \frac{1}{N_1} \sum_{i=1}^{N_1} \big | Z\mathcal{N}_t +  \mu \mathcal{N}_x - g_1 \big |^2, \\
        MSE_{u_0} &= \frac{1}{N_{u_0}} \sum_{i=1}^{N_{u_0}} \big | \mathcal{N} - u_0 \big |^2, \label{eqn: initial displacement loss1D}\\
        MSE_{v_0} &= \frac{1}{N_{v_0}} \sum_{i=1}^{N_{v_0}} \big | \mathcal{N}_t - v_0 \big |^2, \label{eqn: initial velocity loss1D}%
    \end{align}
    \label{eqn: losses1D}%
\end{subequations}
where $N_\Omega$, $N_0$, $N_1$, $N_{u_0}$, $N_{v_0}$ correspond to the number of randomly distributed collocation points in the relevant subdomains, i.e. the domain interior, left boundary ($x = 0$), right boundary ($x = 1$), and at $t = 0$ (for initial displacements and velocities), respectively.  In addition to \eqref{eqn: losses1D} we have the contributions to the total loss from the state evolution equation, namely
\begin{subequations}
    \begin{align}
        MSE_\psi &= \frac{1}{N_\psi}\sum_{i=1}^{N_\psi} \big |\mathcal{N}^\psi_{t} - G(V, \psi) - h\big |^2,\label{eqn: aging loss}\\
        MSE_{\psi_0} &= \frac{1}{N_{\psi_0}}\sum_{i=1}^{N_{\psi_0}} \big |\mathcal{N}^\psi - \psi_0 |^2,\label{eqn: initial state}
    \end{align}
    \label{eqn: aging losses1D}
\end{subequations}
so that the objective function we seek to minimize is given by $MSE = \sum_{\xi\in \chi} MSE_{\xi}$, where subscript $\chi = \{\Omega, 0, 1, u_0, v_0, \psi, \psi_0\}$ so that the sum involves contributions from all subdomains. Minimizing the $MSE$ over the network parameters of $\mathcal{N}$ and $\mathcal{N}^\psi$ define proxy models which produce approximations to $u$ and $\psi$. 

An important detail about the network training involved in minimizing the objective function $MSE$ defined in the previous paragraph is that the inclusion of the state evolution means we are now solving a coupled system of PDE. The coupling occurs at the fault $x=0$, where two conditions (fault friction \eqref{eqn: 1D fault condition2} and state evolution \eqref{eqn: aging}) are now enforced. Because of this coupling, the training networks $\mathcal{N}$ and $\mathcal{N}^\psi$ both appear in the fault loss \eqref{eqn: fault loss1D} as well as in the state evolution loss \eqref{eqn: aging loss}. This interconnectivity may cause training to favor accuracy in the displacement network over accuracy in the state network. In particular, high temporal variations in state evolution, variations in scale, and/or network architecture may drive the training step to favor the displacement approximation. We found it helpful to isolate the state and displacement networks during the backpropogation step. To do this, we define two objective functions: The state objective function consists of loss components in equation \eqref{eqn: aging losses1D} while the displacement objective function uses loss components from equation \eqref{eqn: losses1D}. Each training iteration then requires two optimization steps to update both state and displacement networks. This setup ensures that each network is only updated by one associated objective function. We found this approach to improve our network's training speed and resulted in better approximations of the state variable. The particulars of this training approach are illustrated in Algorithm~\ref{alg: RSF pseudocode} which provides the pseudocode for solving the 1D IBVP given by equations \eqref{eqn: 1Delastodynamic} and \eqref{eqn: 1D boundary conditions} where state evolves according to \eqref{eqn: aging}. 

In order to verify our computational framework for the 1D problem, we generate a known, manufactured solution, which enables direct comparison with the approximate solution produced by the neural network. This approach, known as the method of manufactured solutions \citep{Roache}, assumes a particular analytic solution $u^e$ and derives consistent source terms and boundary data for the IBVP, without changing the underlying PDE and boundary operators. 
We manufacture $u^e$ and $\psi^e$ to be a solution to the IBVP and aging law, respectively, such that 
\begin{align}
    u^e(x, t) &= \tanh{\big (0.5(x - ct + 1)\big ) },\\
    \psi^e(t) &= -\frac{\mu u^e_x (0,t)}{\bar{\sigma}_n} - a\ln \bigg ( \frac{2u^e_t(0,t)}{V_0} \bigg )
\end{align}
This choice of manufactured solution defines the initial data $u_0$ and $v_0$, the source term $s$, and all boundary data, namely 
\begin{subequations}   
    \begin{alignat}{2}
        s &= u^e_{tt} - c^2 u^e_{xx}, \quad && \text{on } \Omega \times [0,T],\\
        g_0 &= 0,\quad &&\text{at }  x=0, \label{eqn: fric_data1D}\\
        g_1 &= Zu^e_t + \mu u^e_x, \quad &&\text{at }  x=1,\\
         h &= \psi^e_t - G(2u^e_t, \psi^e), \quad && \text{at } x=0,
    \end{alignat}
    \label{eqn: 1D source terms}%
\end{subequations}
where, for the 1D problem, $V(t) = 2u_t(0, t)$, and corresponds to a commonly chosen exact solution \cite{Erickson2014, Harvey2022}.  

\begin{figure}[H]
    \centering
    \includegraphics[width=\textwidth]{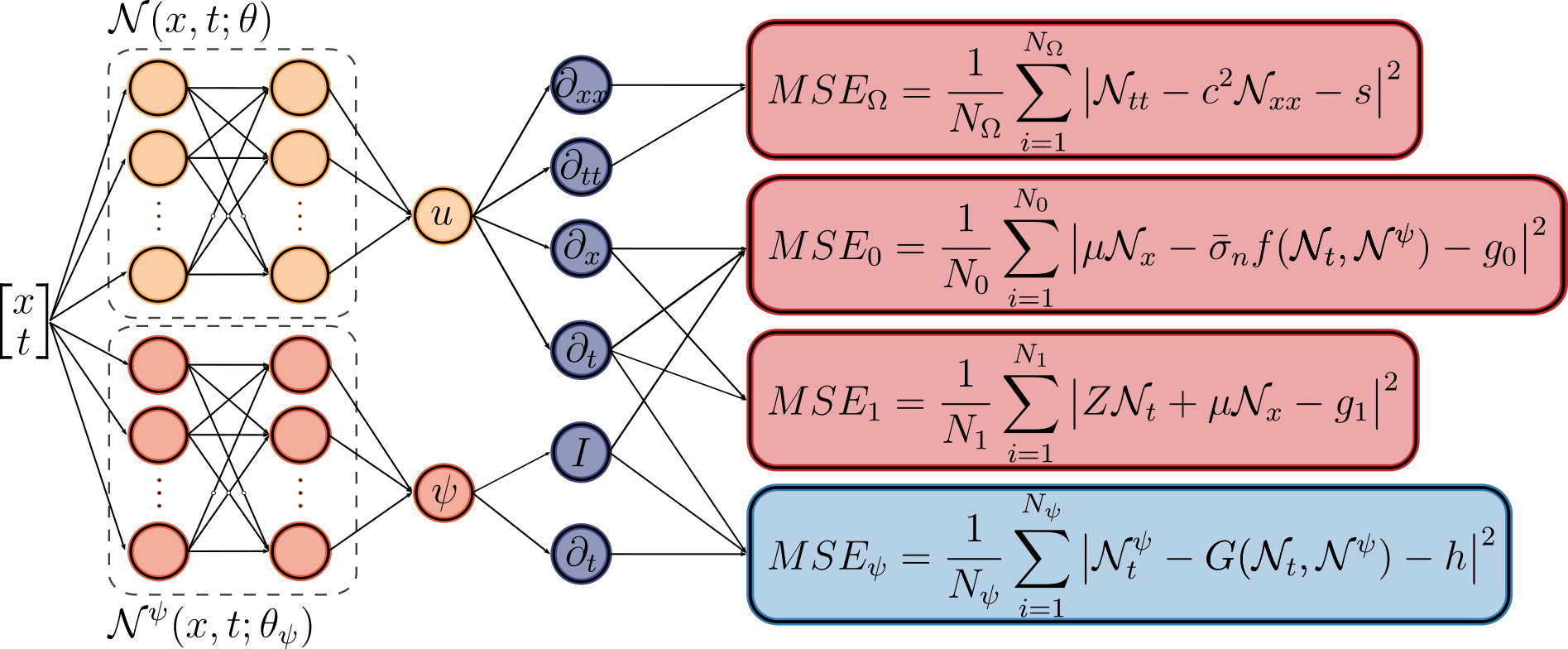}
    \caption{A schematic of the PINN framework for solving the 1D IBVP \eqref{eqn: 1Delastodynamic} \eqref{eqn: 1D boundary conditions} with rate and state friction defined by \eqref{eqn: 1D RNS fritction} and \eqref{eqn: aging}. Displacement network $\mathcal{N}$ and state network $\mathcal{N}^\psi$ are trained separately by defining two objective functions which are differentiated by red and blue nodes, respectively, in the final layer. At each training iteration, $\mathcal{N}$ is updated using component losses $MSE_\Omega, MSE_0,$ and $MSE_1$ while $\mathcal{N}^\psi$ is updated using just $MSE_\psi$ as an objective function.}
    \label{fig: 1D RSF PINN diagram}%
\end{figure}

\begin{figure}[H]
    \centering
    \includegraphics[width=\textwidth]{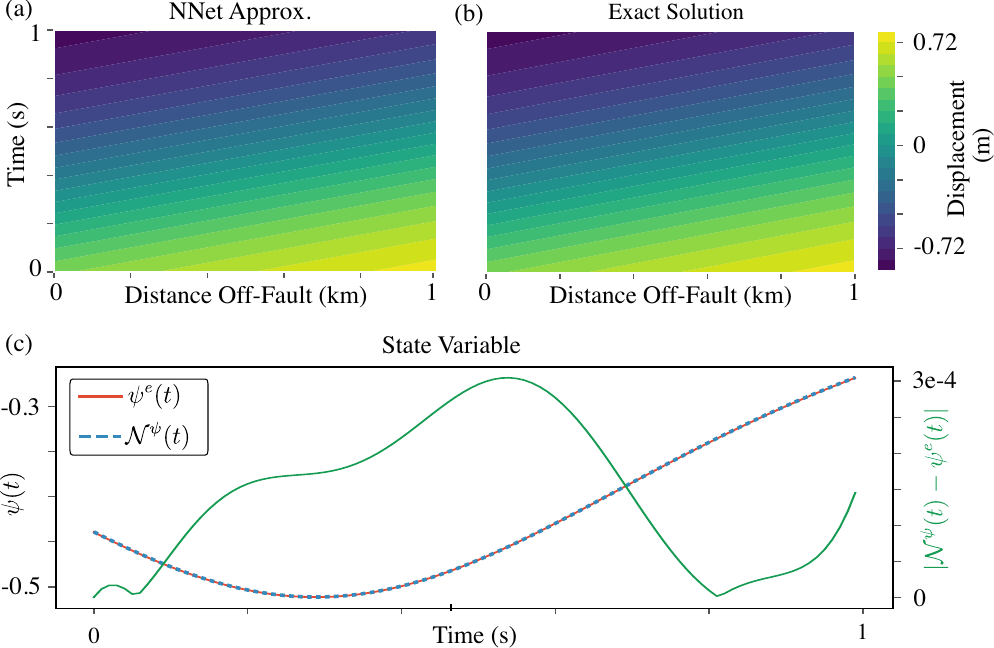}
    \caption{Comparison of results from 1D illustration showing the (a) displacement network approximation $\mathcal{N}$ with (b) manufactured solution $u^e$. Additionally, the (c) state approximation network $\mathcal{N}^\psi$ is plotted against the manufactured state $\psi^e$ along with their absolute error $|\mathcal{N}^\psi(t) - \psi^e(t)|$. Absolute displacement error was averaged over 1000 randomly sampled points and measured to be $|\mathcal{N}-u^e|_{\text{avg}}=1.57e-5$.}
    \label{fig: 1D displacements and state}
\end{figure}

Parameter values used for all studies in this work are given in Table~\ref{tab: mms parameter values}. We provide Figure~\ref{fig: 1D RSF PINN diagram} to illustrate how the independent networks $\mathcal{N}$ and $\mathcal{N}^\psi$ are used to build each component loss function and Algorithm~\ref{alg: RSF pseudocode} provides an outline of the implemented code which is publicly available on GitHub \cite{EQP}.
\begin{algorithm}[t]
\scriptsize
  \begin{algorithmic}[1]
  \State $\mathcal{N} \gets$ \Call{NeuralNetwork}{$x, t; \theta$} \Comment{initialize network parameters $\theta$ for $\mathcal{N}$}
  \State $\mathcal{N}^\psi \gets$ \Call{NeuralNetwork}{$t; \theta_{\psi}$} \Comment{initialize network parameters $\theta_\psi$ for $\mathcal{N}^\psi$}
  \State $\mathcal{N}$opt $\gets$ \Call{SetOptimizer}{$\theta$}  \Comment{Instantiate an optimizer which only tracks parameters of $\mathcal{N}$}
  \State $\mathcal{N}^\psi$opt$ \gets$ \Call{SetOptimizer}{$\theta_\psi$} \Comment{Instantiate an optimizer which only tracks parameters of $\mathcal{N}^\psi$}
  \State
  \State
  \State $\widetilde{\mathcal{N}} \gets u_0(x) + tv_0(x) + t^2 \mathcal{N}(x, t)$ 
  \Comment{Employ hard enforcement of initial conditions}
  \State $\widetilde{\mathcal{N}}^\psi \gets \psi_0 + t\mathcal{N}^\psi(t)$ 
  \State

  \For{$\xi$\textbf{ in} $\chi$} \Comment{Loop over each subdomain and define the relevant component loss}
  \Function{MSE$_\xi$}{$x, t$} 

  \State output $\gets$ \Call{Condition$\big [\xi \big ]$}{$\widetilde{\mathcal{N}}(x, t)$, $\widetilde{\mathcal{N}}^\psi(t)$}
  \Comment{Conditions specified by equations \eqref{eqn: 1Delastodynamic}, \eqref{eqn: 1D boundary conditions},  \eqref{eqn: strength}, and \eqref{eqn: aging}}
  \State data $\gets$ \Call{SourceData$\big [\xi \big ]$}{$x, t$}
  \Comment{Source terms are determined in equation \eqref{eqn: 1D source terms}}
  \State \Return \Call{MeanSquareError}{output, data}
  \EndFunction
  \EndFor
  \State 
  \For {i=1 \textbf{to} training\_iterations} 

  \State Loss$_\mathcal{N} \gets$ 0
  \State Loss$_{\mathcal{N}^\psi} \gets$ 0
  \State \For {$\xi$\textbf{ in }$\chi \setminus \{\Psi \}$}
      \State $x_\xi, t_\xi \gets $\Call{rand}{$N_\xi, \xi$} \Comment{Generate $N_\xi$ randomly sampled points from subdomain $\xi$}
      \State Loss$_\mathcal{N} \gets$ Loss$_\mathcal{N}$ +  \Call{MSE$_\xi$}{$x_\xi, t_\xi$}
      \EndFor
  \State
  \State $x_{\Psi}, t_{\Psi} \gets $\Call{rand}{$N_{\Psi}, {\Psi}$}
  \State Loss$_{\mathcal{N}^\psi} \gets$ \Call{MSE$_{f_\psi}$}{$x_{\Psi}, t_{\Psi}$} 
  \State 
  \State $\nabla_\psi \gets$ grad(Loss$_{\mathcal{N}^\psi}$, [$\theta_\psi$]) \Comment{Compute network gradient}
  \State $\theta_\psi \gets$ $\mathcal{N}^\psi$opt.\Call{Step}{$\nabla_\Psi$} \Comment{Update network weights using the optimization algorithm}
  \State 
  \State $\nabla_\mathcal{N} \gets$ grad(Loss$_\mathcal{N}$, [$\theta$])
  \State $\theta \gets$ $\mathcal{N}$opt.\Call{Step}{$\nabla_\mathcal{N}$} 
  \EndFor
  
 \caption{Multi-network training for 1D forward problem with RSF.}\label{alg: RSF pseudocode}
\end{algorithmic}
\end{algorithm}
Here we take $\mathcal{N}:\mathbb{R}^2 \to \mathbb{R}$ and state variable network $\mathcal{N}^\psi:\mathbb{R} \to \mathbb{R}$ to both be fully-connected, feed-forward networks with three hidden layers, 64 neurons per layer and use hyperbolic tangent activation functions.  The loss function is optimized using L-BFGS\cite{byrd1995limited}, a quasi-Newton optimization algorithm. Minimization is done over 10 training iterations where each iteration is trained on a random sampling of $N = 100$ interior points and 50 boundary points ($N_b = 25$ points per boundary). Network weights are initialized using uniform Xavier initialization \citep{glorot2010understanding}. In Figure~\ref{fig: 1D displacements and state} we show learned approximations for displacement and state compared to their respective exact solutions. These results suggest that our neural network parameters can solve the 1D forward problem to reasonable accuracy. To further explore performance however, and since we are primarily interested in higher dimensional settings that enable the inversion of depth-dependent parameters, we now move to the 2D formulation. 

\subsection{2D Application in Seismic Faulting} 

The 2D problem is obtained by first considering a bounded spatial domain in $\mathbb{R}^3$ defined by $(x, y, z)\in [-L_x, L_x]\times [-L_y, L_y] \times [0, L_z]$ where $z$ is taken to be positive downward. We assume antiplane shear deformation, taking displacements $u_x, u_z$ to be zero, and assume the out-of-plane displacement $u = u_y$ is independent of $y$. Momentum balance for an elastic solid then gives rise to the elastodynamic wave equation in 2D, namely,
\begin{align}
    u_{tt} = c^2 \Delta u + S(x, z, t), \quad (x, z)\in [-L_x, L_x] \times [0, L_z], \,\, t \geq 0,
    \label{eqn: elastodynamic}%
\end{align}
where $\Delta = \left(\dfrac{\partial^2 }{\partial x^2} + \dfrac{\partial^2 }{\partial z^2} \right)$ is the 2D Laplacian. Here the shear wavespeed $c = \sqrt{\mu/\rho}$ as in the 1D case, and $S$ comprises body forces. Assumed initial conditions are given by 
\begin{subequations}
    \begin{align}
        u(x, z, 0) &= u_0(x, z),\\
        u_t(x, z, 0) &= v_0(x, z).
    \end{align}
    \label{eqn: initial conditions}
\end{subequations}
\noindent We assume $z = 0$ corresponds to Earth's surface which we take to be traction free.  An RSF vertical fault is embedded at the interface $x = 0$, corresponding to interface conditions
\begin{subequations}
    \begin{align}
        \tau^+ &= -\tau^-\label{eqn: trac}\\
        \tau^\pm &= F(V^\pm, \psi) \label{eqn: fric}.
    \end{align}
\end{subequations}
Here the traction $\tau$ at a boundary or interface is defined with respect to the outward pointing unit normal $\bf n$ 
by $\tau = {\bf n} \cdot \mu \nabla u$, and $\tau^\pm = \tau(0^\pm, z, t)$. Equation \eqref{eqn: trac} corresponds to the requirement that the traction be ``equal and opposite" across the fault and \eqref{eqn: fric} is a requirement that fault shear stress be equal to frictional strength $F$ as in the 1D case. As in the 1D illustration, $F$ is a nonlinear function of the slip rate $V^\pm(z, t) = \dot{u}(0^\pm, z, t)-\dot{u}(0^{\mp},z,t)$ and an empirical state variable $\psi(z, t)$ the particular form of which is given by \eqref{eqn: strength}.  At the remote boundaries $(x, z) = (\pm L_x, L_z)$ we assume the boundaries are non-reflecting. Next we make the assumption that the displacement field $u(x,z,t)$ is antisymmetric about the fault interface $x=0$, which automatically satisfies \eqref{eqn: trac},  so that we may restrict our focus to one side of the fault as in \citet{Erickson2014} and reduce the computational redundancy.  Our final spatial domain is then $\Omega = [0, L_x]\times [0, L_z]$ where the fault interface now becomes a boundary.  For generality (and to aid in later verification steps) we state the final boundary conditions on $\partial\Omega$ using generic data $g_f, g_s, g_r, g_d$ for the fault, surface, remote, and depth boundaries, respectively, namely
\begin{subequations}
    \begin{alignat}{2}
        \tau &= F(V) + g_f(z,t), \quad &&x=0, \label{eqn: fault condition}\\
        \tau &= g_s(x,t), \quad &&z=0, \\
        Zu_t + \tau &= g_r(z,t), \quad &&x=L_x, \\
        Zu_t + \tau &= g_d(z,t), \quad &&z=L_z,
    \end{alignat}
    \label{eqn: boundary conditions}%
\end{subequations}
although the Earth's free surface and non-reflecting conditions are the far field boundaries correspond to $g_s = g_r = g_d = 0$.  Here $\tau = \tau^+$, $V = V^+ = 2u_t^+$ and $Z=\sqrt{\mu \rho}$ is the shear impedance as in the 1D case.


%
%

For this 2D application problem we are primarily interested in the RSF parameter $a - b$, which describes the velocity-dependence of friction at steady state. Positive values (i.e. $a > b$) correspond to stable sliding, while negative values correspond to frictional instabilities. Knowledge of the depth-distribution of $a-b$ is therefore of central importance to understanding slip behavior and deformation \citep{Brodsky2020}.  In other words, heterogeneities along fault interfaces can be characterized at least in part by velocity-weakening frictional behavior ($a-b < 0$) indicating that seismic rupture may nucleate and easily propagate, while stable regions are characterized by velocity-strengthening frictional behavior ($a - b > 0$) that inhibits the sliding at seismogenic speeds \citep{scholz_2019}.  In order to focus our study on the inference of the depth-dependency of $a-b$, we assume that the state variable is at steady state, namely $\psi = f_0 + b \ln\left(\frac{V_0}{V}\right)$, so that the friction coefficient reduces to the rate-dependent form
\begin{equation}
    f(V) = f_0 + \alpha \ln\left(\frac{V}{V_0}\right),
\end{equation}
where we have introduced $\alpha = a-b$. We assume that $\alpha$ is piecewise linear with depth, defining a shallow seismogenic zone, namely,   
\begin{align}
    \alpha (z) = \begin{cases}
        \alpha_\text{min} & 0<z<H\\
        (z - H) * ((\alpha_\text{max}-\alpha_\text{min})/D) + \alpha_\text{min} & H\leq z \leq H+D\\
        \alpha_\text{max} & H+D<z,
    \end{cases}
    \label{eqn: depth dep friction param}%
\end{align}
where $H$ defines the seismogenic depth, $\alpha_\text{min}$ and $\alpha_\text{max}$ are constants defining the minimum and maximum values assumed by $\alpha$, and $D$ the transition distance. In this work we consider PINN solutions to both the forward and inverse problems.  For forward problem we solve for the unknown displacements $u$ in the IBVP \eqref{eqn: ibvp(operator form)} assuming $\alpha(z)$ has been specified.  For the inverse problems, in addition to solving for $u$ we also infer the friction parameter $\alpha$. In this latter case, \eqref{eqn: depth dep friction param} sets the data for the boundary conditions, but $\alpha$ is not assumed to be known a priori when enforcing the frictional interface condition, as will be described in the next section.

\subsection{Forward and Inverse Problems for the 2D Application}
Governing equation \eqref{eqn: elastodynamic} along with initial and boundary conditions \eqref{eqn: initial conditions}, \eqref{eqn: boundary conditions} provide specifics of the loss terms \eqref{eqn: MSE loss} that define the PINN.  However, there is more than one way to formulate the associated learning problem. First, one can consider either a forward or inverse problem \citep{raissi2019physics}. As this work is concerned with generating approximations to both forward and inverse problems, we will refer to the \textit{primal} solution as the network approximation $\mathcal{N}$ to the displacement $u$ specified by the IBVP \eqref{eqn: ibvp(operator form)}. The \textit{forward} problem is only concerned with approximating a primal solution and requires that all model parameters $\lambda$ are known a priori. A solution to the \textit{inverse} problem aims to approximate the primal solution while also being tasked with learning a set of system parameters. If the $\lambda$ in \eqref{eqn: ibvp(operator form)} are known, the forward problem is solved and the problem reduces to an unsupervised learning task \citep{cuomo2022scientific}. In the inverse problem, however, the PINN has the same loss function \eqref{eqn: MSE loss} but with minor changes: Instead of knowing the system parameters $\lambda$, we establish them as trainable networks, which we detail shortly.  

In addition to specifying whether a forward or inverse problem is being solved, in either case one must also choose between soft or hard enforcement of initial conditions \citep{lagaris1998artificial, lagaris2000neural, karniadakis2021physics, cuomo2022scientific}. Soft enforcement uses loss terms to learn boundary data whereas hard enforcement encodes boundary data into a trial function to satisfy the conditions exactly. We proceed with details concerning each of these below.

\subsubsection{Forward Problem}
As in the 1D case, we begin by supposing that $\mathcal{N}(x, z, t; \theta) \approx u(x, z, t)$ for neural network $\mathcal{N}$ and solution $u$ of the IBVP \eqref{eqn: elastodynamic}, \eqref{eqn: initial conditions}, \eqref{eqn: boundary conditions}. The 2D component loss functions are therefore given by
\begin{subequations}
    \begin{align}
        MSE_\Omega &= \frac{1}{N_\Omega}\sum_{i=1}^{N_\Omega} \big |\mathcal{N}_{tt} - c^2\Delta \mathcal{N} -S\big |^2, \\
        MSE_f &= \frac{1}{N_f} \sum_{i=1}^{N_f} \big | -\mu \mathcal{N}_x - \bar{\sigma}_n f - g_f \big |^2, \label{eqn: fault loss}\\
        MSE_s &= \frac{1}{N_s} \sum_{i=1}^{N_s} \big | -\mu \mathcal{N}_z - g_s \big |^2, \\
        MSE_r &= \frac{1}{N_r} \sum_{i=1}^{N_r} \big | Z\mathcal{N}_t +  \mu \mathcal{N}_x - g_r \big |^2, \\
        MSE_d &= \frac{1}{N_d} \sum_{i=1}^{N_d} \big | Z\mathcal{N}_t +  \mu \mathcal{N}_z - g_d \big |^2  \\
        MSE_{u_0} &= \frac{1}{N_{u_0}} \sum_{i=1}^{N_{u_0}} \big | \mathcal{N} - u_0 \big |^2, \label{eqn: initial displacement loss}\\
        MSE_{v_0} &= \frac{1}{N_{v_0}} \sum_{i=1}^{N_{v_0}} \big | \mathcal{N}_t - v_0 \big |^2, \label{eqn: initial velocity loss}%
    \end{align}
    \label{eqn: antiplane losses}%
\end{subequations}
where, as in the 1D case, summations are over randomly distributed collocation points in the interior and domain boundaries. The 2D objective function we seek to minimize is given by $MSE = \sum_{\xi\in \chi} MSE_{\xi}$, where (as in the 1D case),  subscript $\chi = \{\Omega, f, s, r, d, u_0, v_0\}$ implies that the sum is taken over all relevant subdomains.

\subsubsection{Inverse Problem}
\begin{figure}[H]
    \centering
    \includegraphics[width=\textwidth]{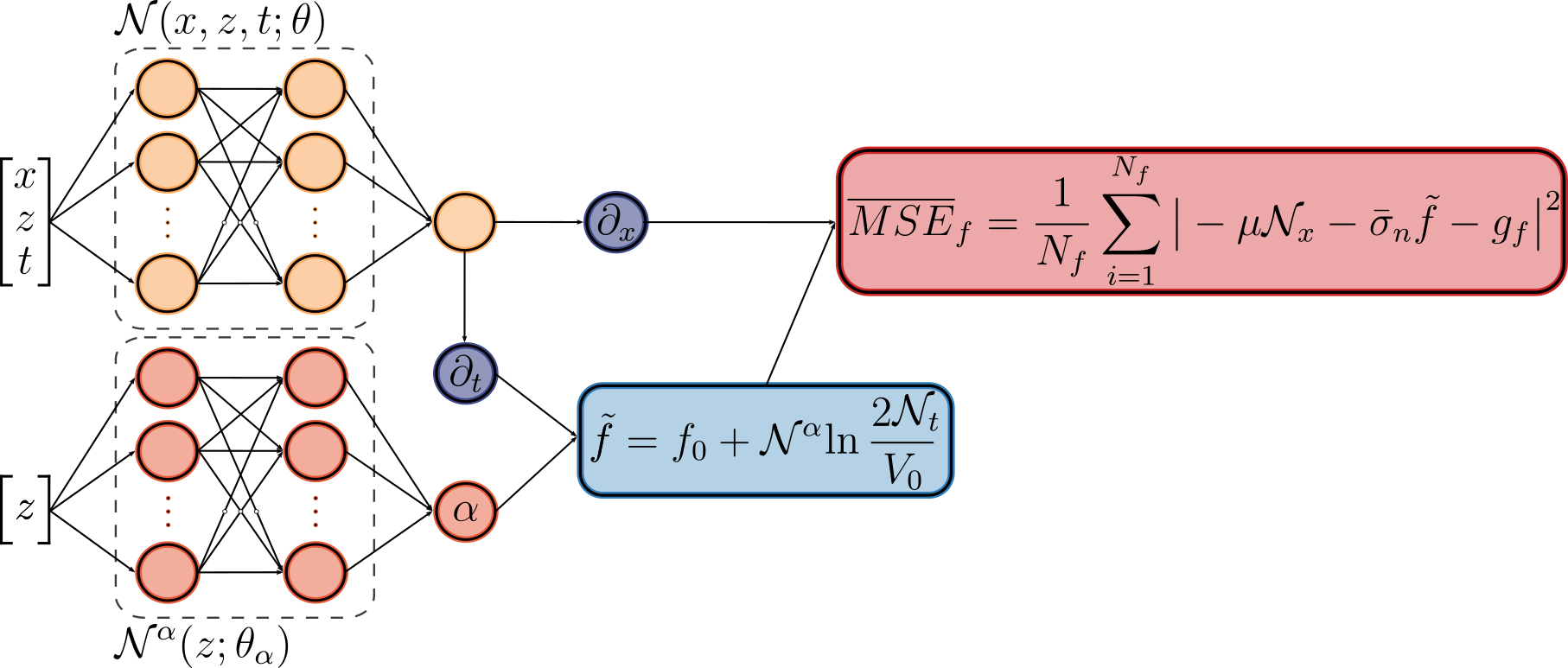}
    \caption{Network diagram for enforcing the stress condition along the fault while also learning the depth-dependent friction parameter $\alpha$. During training, the $z$-coordinate of a training point is passed to both $\mathcal{N}$ and $\mathcal{N}^\alpha$.}
    \label{fig: fault network}%
\end{figure}

While the primal solution to an IBVP is desirable, replacing an assumed model parameter (which often comes with much uncertainty) with a trainable network could leverage real-world data to physically constrain parameter distributions. The inverse problem is derived by making a slight modification to the forward problem \eqref{eqn: antiplane losses}. As we are interested in learning the fault friction parameter $\alpha$, we modify the loss component \eqref{eqn: fault loss} to include a trainable network. Because $\alpha$ is depth-variable it must be explicitly trained on random collocation points along the fault. So in addition to the first network $\mathcal{N}(x, z, t; \theta) \approx u(x, z, t)$, we introduce a secondary network $\mathcal{N}^\alpha(z; \theta_\alpha) \approx \alpha(z)$. 
Only one loss component needs be changed from the forward problem \eqref{eqn: antiplane losses}, namely, component \eqref{eqn: fault loss}, which enforces the fault friction condition. A modified friction coefficient is considered, defined by
\begin{align}
    \tilde{f} = f_0 + \mathcal{N}^\alpha(z) \ln{\bigg [\frac{2\mathcal{N}_t(z,t)}{V_0} \bigg ]},
    \label{eqn: inferred friction}%
\end{align}
which gives rise to the modified fault loss
\begin{align}
    \overline{MSE}_f &= \frac{1}{N_f} \sum_{i=1}^{N_f} \big | -\mu \mathcal{N}_x - \bar{\sigma}_n \tilde{f} - g_f \big |^2.
    \label{eqn: modified fault loss}%
\end{align}
Replacing loss component \eqref{eqn: fault loss} in the forward problem with \eqref{eqn: modified fault loss} defines the inverse problem. We give a diagram of the fault network for the inverse problem in Figure~\ref{fig: fault network} showing how the fault loss is used to update two networks during training. The inverse problem is trained over the same subdomains as the forward problem \eqref{eqn: antiplane losses} but the depth coordinate for fault training data is passed to both $\mathcal{N}$ and $\mathcal{N}^\alpha$.

\subsubsection{Soft vs. Hard Enforcement of Boundary Conditions}
Both the forward and inverse problems require that we specify initial and boundary conditions, which may be enforced in two possible ways. \textit{Soft} enforcement is done by penalizing the objective function, as the network output need not satisfy the condition exactly (as was done in the 1D example of the previous section). Up until now, we have presented soft enforcement, as given in \eqref{eqn: antiplane losses}, which shows component losses corresponding to the boundary and initial conditions. Soft enforcement does not give us any guarantee regarding accuracy of the condition being enforced and each additional loss term in the objective function increases the complexity of the optimization landscape. 

Alternatively, we may enforce initial/boundary conditions in a manner that reduces the complexity of the network training space by encoding such conditions into the network architecture. This approach, known as \textit{hard} enforcement, requires that we specify a trial function which automatically satisfies the initial/boundary conditions. Here we apply this technique to enforce initial conditions, which reduces the number of constraints on the system. Let $\mathcal{N}({\bf x}, t; \theta)$ be a feed-forward neural network and define the trial function $\mathcal{\widetilde{N}}$ to be
\begin{align}
    \mathcal{\widetilde{N}}({\bf x}, t; \theta) = u_0({\bf x}) + tv_0({\bf x}) + t^2 \mathcal{N}({\bf x}, t; \theta)
    \label{eqn: trial function}%
\end{align}
where ${\bf x} = (x, z)$, so that $\mathcal{\widetilde{N}}$ satisfies both initial conditions \eqref{eqn: initial conditions} exactly and is trained by training the network $\mathcal{N}$. If $\mathcal{\widetilde{N}}$ is used in place of $\mathcal{N}$ in the formulation of the objective function, there is no longer a need to include loss terms for initial displacements \eqref{eqn: initial displacement loss}, or initial velocities \eqref{eqn: initial velocity loss} so the network can be trained on a less restrictive set of conditions.

\section{2D Verification, Validation and Applications}\label{sec: verification}

When computational methods for physical problems are used to address science questions, verification is an essential first step to ensure credible results \citep{Harris2009, Erickson2020}. While validation with observational data is the focus of future work, we must first verify that our physics-informed deep learning framework is able to solve both forward and inverse problems to reasonable accuracy.  

%
\begin{table}[t]
    \centering
    \begin{tabular}{c c}
        \hline
        Parameter & Value  \\
        \hline
        $L_x, L_z$ & 25 \si{\kilo \meter}\\
        $H$ & 12 \si{\kilo \meter}\\
        $D$ & 5 \si{\kilo \meter}\\
        $\mu$  & 32 \si{\giga \pascal} \\
        $\rho$ & 2.67 \si[per-mode = symbol]{\kilo \gram \per \meter^3 }\\
        $\alpha_\text{min}$ & -0.005\\
        $\alpha_\text{max}$ & 0.015\\
        $f_0$  & 0.6\\
        $\bar{\sigma}_n$ & 50 \si{\mega\pascal}  \\
        $V_0$  & $10^{-6}$ \si[per-mode = symbol]{\meter \per \second}\\
        $D_c$ & 2 \si{\meter}
    \end{tabular}
    \caption{Parameter values used in the manufactured solution tests.}
    \label{tab: mms parameter values}%
\end{table}

\subsection{Verification with the Method of Manufactured Solutions}
As in the 1D case, we verify the PINN framework by generating a known, manufactured solution $u^e$ which solves the IBVP \eqref{eqn: ibvp(operator form)}, \eqref{eqn: initial conditions}, \eqref{eqn: boundary conditions}. We take 
\begin{align}
    u^e(x,z,t) = \tanh{\big ((x + z + ct)/20 \big )},
\end{align}
which defines the initial data $u_0$ and $v_0$, the source term $S$, and all boundary data, which for the forward problem is
\begin{subequations}   
    \begin{alignat}{2}
        S &= u^e_{tt} - c^2 \Delta u^e, \quad && \text{on } \Omega \times [0,T],\\
        g_f &= -\mu u^e_{x} - \bar{\sigma}_n f \big ( 2u^e_t \big ),\quad &&\text{at }  x=0, \label{eqn: fric_data}\\
        g_s &= -\mu u^e_{z}, \quad &&\text{at }  z=0,\\
        g_r &= Zu^e_t + \mu u^e_x, \quad &&\text{at }  x=L_x,\\
        g_d &= Zu^e_t + \mu u^e_z,  \quad &&\text{at }  z=L_z.
    \end{alignat}
    \label{eqn: source terms}%
\end{subequations}
The parameters used are slightly modified from the earthquake simulations of \citet{Harvey2022}, given in Table~\ref{tab: mms parameter values}, and we default to this parameter set unless stated otherwise. Note that in the case of the inverse problem, $\alpha_\text{min}$ and $\alpha_\text{max}$ define the manufactured exact solution $\alpha^e(z)$, which is used only to set the data in \eqref{eqn: fric_data}; $\alpha(z)$ is a learned parameter through the use of the network $\mathcal{N}^\alpha$. 
In all scenarios we consider, we take the primal network $\mathcal{N}:\mathbb{R}^3 \to \mathbb{R}$ and friction network $\mathcal{N}^\alpha:\mathbb{R} \to \mathbb{R}$ to both be fully-connected, feed-forward networks with three hidden layers and 128 neurons per layer. The wave equation \eqref{eqn: elastodynamic} requires displacements to be sufficiently smooth with respect to the input but no such requirement is imposed on the friction parameters. Thus we use hyperbolic tangent activation functions for the displacement network but use linear units in the friction network. We found that the friction network performed well when we used both Rectified Linear Unit (ReLU) and Sigmoid Linear Unit (SiLU) together. We use ReLU activation for the outer hidden layers and a SiLU activation on the interior layer.  
The code used to perform the numerical simulations in this work is developed in Python using Pytorch and is publicly available \cite{EQP}. Additionally, we provide Algorithm~\ref{alg: pseudocode} as a sketch of the implemented code used to solve the inverse problem with hard enforcement of initial conditions.
\begin{algorithm}[t]
\scriptsize
  \begin{algorithmic}[1]
  \State $\mathcal{N} \gets$ \Call{NeuralNetwork}{$x, z, t; \theta$} \Comment{$\theta$ initial network parameters for $\mathcal{N}$}
  \State $\mathcal{N}^\alpha \gets$ \Call{NeuralNetwork}{$z; \theta_{\alpha}$} \Comment{$\theta_\alpha$ initial network parameters for $\mathcal{N}^\alpha$}
  \State $\theta \gets [\theta, \theta_\alpha]$
  \State
  \State $\widetilde{\mathcal{N}} \gets u_0(x,z) + tv_0(x, z) + t^2 \mathcal{N}(x, z, t)$ \Comment{Employ hard enforcement of initial conditions}
  \State

  \For{$\xi$\textbf{ in }$\chi$} \Comment{Loop over each subdomain and define the relevant component loss}
  \Function{MSE$_\xi$}{x, z, t} 
  \State output $\gets$ \Call{Condition$\big [\xi \big ]$}{$\widetilde{\mathcal{N}}(x, z, t)$, $\mathcal{N}^\alpha(z)$}
  \State data $\gets$ \Call{SourceData$\big [\xi \big ]$}{$x, z, t$}

  \State \Return \Call{MeanSquareError}{output, data}
  \EndFunction
  \EndFor
  \State 
  \For {i=1 \textbf{to} training\_iterations} 

  \State Loss $\gets$ 0
  \State \For {$\xi$ \textbf{in} $\chi$}
      \State ${\bf x}_\xi \gets $\Call{rand}{$N_\xi, \xi$} \Comment{Generate $N_\xi$ randomly sampled points from subdomain $\xi$}
      \State Loss $\gets$ Loss +  \Call{$MSE_\xi$}{${\bf x}_\xi $}
      \EndFor
  \State
  \State $\nabla \gets$ grad(Loss, [$\theta$])
  \State $\theta \gets$ \Call{OptimizerStep}{$\nabla$} \Comment{Update network weights using the optimization algorithm}
  \EndFor
  
 \caption{Multi-network training for solving the inverse problem}\label{alg: pseudocode}
\end{algorithmic}
\end{algorithm}

$\mathcal{N}$ and (in the case of the inverse problem) $\mathcal{N}^\alpha$ are trained by minimizing the MSE (the sum of component MSE given in \eqref{eqn: antiplane losses})  where all loss functions are optimized using L-BFGS \cite{byrd1995limited}, a quasi-Newton optimization algorithm. Minimization is done over 30 training iterations where each iteration is trained on a random sampling of $N = 400$ interior points and 400 boundary points ($N_b = 100$ points per boundary). Additionally, if soft enforcement of initial conditions is used, we sample 400 additional interior points (200 for displacements and 200 for velocities). After the 30 training iterations are complete we evaluate each of the component loss functions using 1000 randomly sampled points. Likewise we generate the same number of sample points for $\mathcal{N}$ and $\mathcal{N}^\alpha$ and use their respective manufactured solutions to compute the relative $\ell^2-$norm of their errors, namely, 
\begin{align}
    \norm{\mathcal{N} - u^e}^2_{\text{1, rel}} &= \frac{\sum_{i=1}^N |\mathcal{N}(x_i, z_i, t_i) - u^e(x_i, z_i, t_i)|^2}{\sum_{i=1}^N |u^e(x_i, z_i, t_i)|^2},\\
    \norm{\mathcal{N_\alpha} - \alpha^e}^2_\text{1, rel} &= \frac{\sum_{i=1}^{N_b} |\mathcal{N}^\alpha(z_i) - \alpha^e(z_i)|^2}{\sum_{i=1}^{N_b} |\alpha^e(z_i)|^2}.
\end{align}

Network weights are initialized using uniform Xavier initialization \citep{glorot2010understanding}. Variation due to weight initialization involving randomization is accounted for by averaging error data across $50$ trained solutions. Table \ref{tab:losses and errors} presents error data for the forward and inverse problems where both hard and soft enforcement of initial conditions are considered for each. As the table illustrates, for both the forward and inverse problems, hard enforcement of initial conditions are accompanied with a smaller $\ell^2$-errors after 30 training iterations, which is most likely due at least in part to the fact that hard enforcement means that the network approximation satisfies the initial conditions exactly. Also illustrated in the table, is that all component mean-square errors (MSE) are lower for the soft enforcement of initial conditions compared to hard enforcement. Soft enforcement uses two additional loss components (i.e. those enforcing initial conditions) which results in a more biased model, or a lower variance model (i.e. the MSE is less sensitive to training data). The trade off between bias and variance (introduced via randomness in weight initialization and collocation points) may account for such lower MSE when using soft enforcement (which involves lower variances) and averaging over $50$ trained models.

\begin{table}[t]
    \centering
    \renewcommand{\arraystretch}{1.5}
    \scriptsize
    \begin{tabularx}{\textwidth}{X X X X X X X}
        \toprule
        Errors   & \multicolumn{3}{l}{Forward Problem}   & \multicolumn{3}{l}{Inverse Problem} \\
        \cmidrule(lr){2-4}\cmidrule(l){5-7}
        & Soft & Hard &Error Ratio  &Soft  &Hard  &Error Ratio \\
        \midrule
        $\norm{\mathcal{N} - u^e}_\text{1, rel}$  &1.364e-02 & 7.148e-03 &1.908e+00  &2.931e-02 & 1.633e-02 &1.795e+00\\
        $\norm{\mathcal{N_\alpha} - \alpha^e}_\text{1, rel}$  &  &  &  &  2.977e-01 & 7.220e-02 &4.123e+00\\
        $MSE_{PDE}$  &1.149e-03 & 8.883e-03 &1.293e-01  &1.133e-03 & 6.675e-03 & 1.697e-01\\
        $MSE_{f}$    &2.633e-03 & 4.115e-02 &6.399e-02  &8.579e-03 & 1.615e-01 & 5.311e-02\\
        $MSE_{s}$    &8.393e-05 & 7.868e-04 &1.067e-01  &1.251e-04 & 6.545e-04 & 1.911e-01\\
        $MSE_{d}$    &8.216e-05 & 2.115e-03 &3.885e-02  &4.681e-04 & 7.199e-03 & 6.502e-02\\
        $MSE_{r}$    &1.195e-05 & 3.233e-04 &3.698e-02  &1.064e-04 & 8.074e-04 & 1.318e-01\\
        $MSE_{u_0}$  &1.703e-04 &           &  &3.722e-04 &           &\\
        $MSE_{v_0}$  &6.308e-04 &           &  &1.375e-03 &           &\\
        \bottomrule
    \end{tabularx}
    \caption{Errors in the $\ell^2$-norm and MSE (loss) associated with hard and soft enforcement of initial conditions for both the forward and inverse problem. Each term is computed over 1000 randomly sampled points and averaged across 50 trained solutions. Error ratios are computed by dividing soft enforcement error by the hard enforcement error.}
    \label{tab:losses and errors}%
\end{table}

Next we test network performance across various subdomains and resolutions. Let $\mathcal{N}$ be a displacement network trained to solve the inverse problem using hard enforcement of initial conditions. This time, errors are measured with respect to the continuous $L^2$-norm
\begin{align}
    \norm{\mathcal{N}-u^e}^2_{2, \Lambda} = \int_\Lambda |\mathcal{N}(x, z, t) - u^e(x, z, t)|^2 \, d\lambda,\label{eqn: cont L2-err}
\end{align}
and we consider error accumulation across $\Omega$, $[0,T]$ and $\hat{\Omega} = \Omega \times [0, T]$, the spatial, temporal and space-time domains, respectively. The integral in \eqref{eqn: cont L2-err} is approximated using Simpson's composite quadrature rule which converges to the exact integral with rate $\mathcal{O}(k^4)$, where $k$ is the quadrature grid size (i.e. the subinterval length). To account for variations in the trained network, we average errors over points outside of the domain of integration. For example, the average temporal error given by  
\begin{align}
    \overline{\norm{\mathcal{N} - u^e}}_{2, [0,T]} = \frac{1}{N_T}\sum_{k=1}^{N_t} \bigg [ \int_{[0,T]} \big |\mathcal{N}(x_k, z_k, t) - u^e(x_k, z_k, t) \big |^2 \, d\lambda(t) \bigg ]^{1/2}
\end{align}
measures expected error accumulation (over time) given $N_T$ randomly sampled spatial points. 

\begin{figure}[H]
    \centering
    \includegraphics[width=\textwidth]{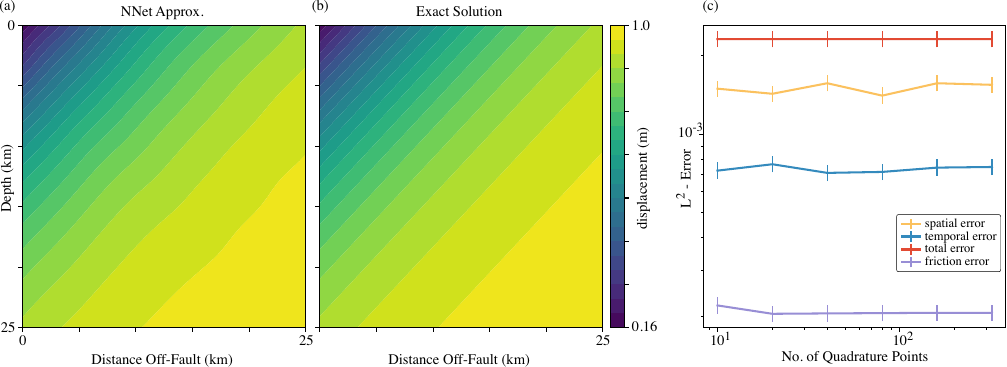}
    \caption{(a) 2D displacement plot for a PINN trained to solve the inverse problem using hard enforcement of initial conditions compared to (b) the manufactured displacements. (c)$L^2-$errors for displacement in space, time, and spacetime (along with $L^2-$errors for the friction parameter) are computed on a uniform grid using Simpson's rule as a quadrature. Errors are then recorded over several mesh refinements.}
    \label{fig: nnet_disp_errs}
\end{figure}

\begin{figure}[H]
    \centering
    \includegraphics[width=\textwidth]{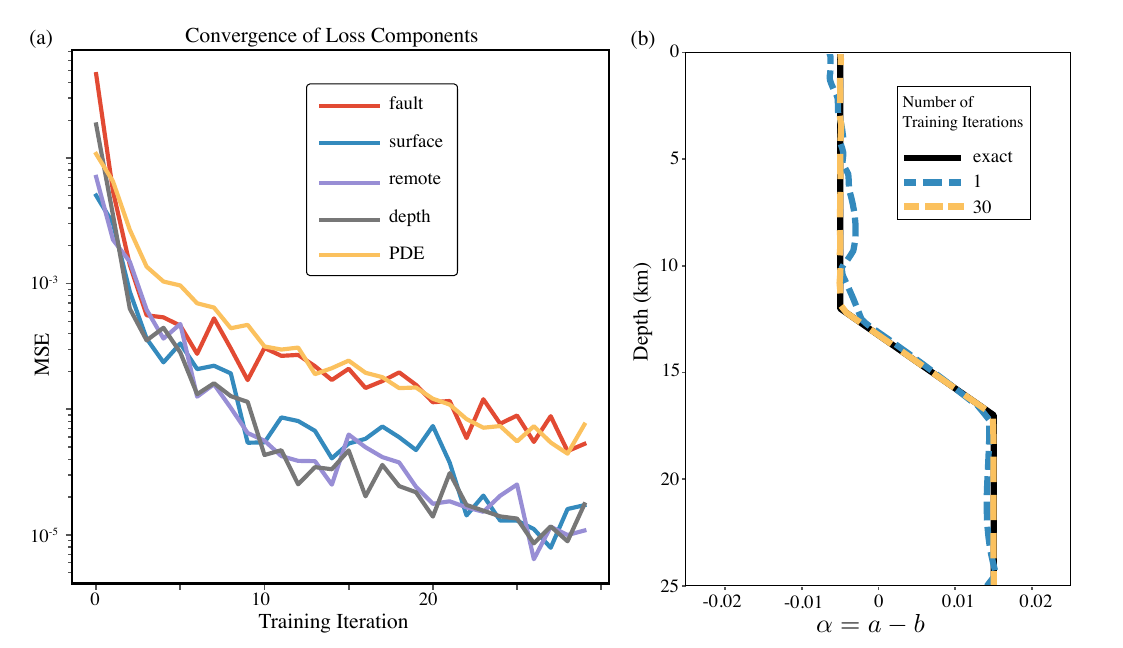}
    \caption{\small 2D inversion results showing (a) convergence of loss components and (b) convergence of the inferred parameter approximation.}
    \label{fig: nnet_fric}
\end{figure}

Figure~\ref{fig: nnet_disp_errs} shows displacement plots for the trained network and the exact solution at final time $T=1$ in addition to network $L^2$-error approximations for displacement in space, time and space-time (along with an $L^2$-error approximation of the friction parameter) for increasingly higher grid resolution used in the numerical quadrature. The errors remain relatively constant with decreasing subinterval size, suggesting that the quadrature approximation is approaching the actual $L^2$-error, and that the PINN approximation maintains good accuracy even when evaluated on higher-resolution grids.  Figure \ref{fig: nnet_fric}(a) shows network component loss functions against training iteration, revealing a non-monotonic decrease across all components. Figure \ref{fig: nnet_fric}(b) illustrates convergence of the inferred parameter $\alpha = a-b$ to the exact distribution $\alpha^e$. Here we show iterations 1 and 30 merely because after the first iteration the inferred parameter performs well and does not vary significantly.

\section{Summary and Future Work}\label{sec: summary}
We have presented a computational framework for physics-informed neural networks (PINNs) for solving the elastodynamic wave equation with a rate-and-state frictional fault boundary in both 1D and 2D. We consider both forward and inverse problems, with the latter obtained by extending to a multi-network architecture in order to learn depth-dependent friction parameters alongside deformations in the Earth's crust. We verified the computational framework by applying the method of manufactured solutions to probe various error measurements. We show that in general, hard enforcement of boundary conditions result in trained networks that better approximate displacements and friction parameters but tend to be worse at minimizing component loss functions when compared to soft enforcement. We found that a PINN defined by hard enforcement of initial conditions produces reasonable approximations for displacements and the desired friction parameter distribution. Though the network is mesh-free, we show that successive mesh refinements of the quadrature approximation to the $L^2$-error yield near constant values, suggesting that the PINN provides a reasonable approximation to the solution even when evaluated on increasingly finer grids. Moreover, although the network requires sufficient training iterations to properly learn displacements, the desired friction parameter is learned within the first couple of iterations. This suggests that PINNs may be a highly effective tool in inferring subsurface friction properties along faults, constrained by both physics and observational surface data. 


While the PINN is shown to perform well when learning the state variable in 1D, and inferring depth-dependency of RSF parameter $a-b$ at steady-state in 2D, we plan to explore the capabilities of the 2D framework with non-steady-state-evolution. This extension requires additional networks in the 2D setting in order to approximate the state variable, and two inference networks to capture the empirical parameters $a$ and $b$ (which become separated across governing equations), and/or networks to learn other frictional parameters, such as $D_c$, whose scaling from laboratory values to actual fault zones is the subject of many studies \cite[e.g.][]{Marone1993}. Approaching such a problem may be aided by a better understanding of the PINN dependence on problem configuration as well as network architecture. Additionally, it would be worthwhile to investigate methods which hybridize PINNs with traditional numerical methods similar to the discrete time PINN in \citet{raissi2019physics}. And finally, with a PINN solution that can handle learning full rate-and-state fault friction in 2D (and eventually 3D), we would be ready to compare model outcomes against community benchmark problems concerning dynamic rupture simulations \cite{Harris2009} and sequences of earthquakes and aseismic slip \cite{Erickson2020}.

\newpage

\let\url\nolinkurl
\bibliographystyle{elsarticle-harv.bst}
\bibliography{main}

\begin{thebibliography}{53}
\expandafter\ifx\csname natexlab\endcsname\relax\def\natexlab#1{#1}\fi
\providecommand{\url}[1]{\texttt{#1}}
\providecommand{\href}[2]{#2}
\providecommand{\path}[1]{#1}
\providecommand{\DOIprefix}{doi:}
\providecommand{\ArXivprefix}{arXiv:}
\providecommand{\URLprefix}{URL: }
\providecommand{\Pubmedprefix}{pmid:}
\providecommand{\doi}[1]{\href{http://dx.doi.org/#1}{\path{#1}}}
\providecommand{\Pubmed}[1]{\href{pmid:#1}{\path{#1}}}
\providecommand{\bibinfo}[2]{#2}
\ifx\xfnm\relax \def\xfnm[#1]{\unskip,\space#1}\fi
\bibitem[{Atwater et~al.(2005)Atwater, Satoko, Kenji, Yoshinobu, Kazue and
  Yamaguchi}]{Atwater2005}
\bibinfo{author}{Atwater, B.F.}, \bibinfo{author}{Satoko, M.R.},
  \bibinfo{author}{Kenji, S.}, \bibinfo{author}{Yoshinobu, T.},
  \bibinfo{author}{Kazue, U.}, \bibinfo{author}{Yamaguchi, D.K.},
  \bibinfo{year}{2005}.
\newblock \bibinfo{title}{The Orphan Tsunami of 1700: Japanese Clues to a
  Parent Earthquake in North America}.
\newblock \bibinfo{edition}{2} ed., \bibinfo{publisher}{University of
  Washington Press}.
\newblock \URLprefix \url{http://www.jstor.org/stable/j.ctvcwnbrv}.
\bibitem[{Bergen et~al.(2019)Bergen, Johnson, de~Hoop and Beroza}]{Bergen2019}
\bibinfo{author}{Bergen, K.J.}, \bibinfo{author}{Johnson, P.A.},
  \bibinfo{author}{de~Hoop, M.V.}, \bibinfo{author}{Beroza, G.C.},
  \bibinfo{year}{2019}.
\newblock \bibinfo{title}{Machine learning for data-driven discovery in solid
  earth geoscience}.
\newblock \bibinfo{journal}{Science} \bibinfo{volume}{363},
  \bibinfo{pages}{eaau0323}.
\newblock \DOIprefix\doi{10.1126/science.aau0323}.
\bibitem[{Brodsky et~al.(2020)Brodsky, Mori, Anderson, Chester, Conin, Dunham,
  Eguchi, Fulton, Hino, Hirose, Ikari, Ishikawa, Jeppson, Kano, Kirkpatrick,
  Kodaira, Lin, Nakamura, Rabinowitz, Regalla, Remitti, Rowe, Saffer, Saito,
  Sample, Sanada, Savage, Sun, Toczko, Ujiie, Wolfson-Schwehr and
  Yang}]{Brodsky2020}
\bibinfo{author}{Brodsky, E.E.}, \bibinfo{author}{Mori, J.J.},
  \bibinfo{author}{Anderson, L.}, \bibinfo{author}{Chester, F.M.},
  \bibinfo{author}{Conin, M.}, \bibinfo{author}{Dunham, E.M.},
  \bibinfo{author}{Eguchi, N.}, \bibinfo{author}{Fulton, P.M.},
  \bibinfo{author}{Hino, R.}, \bibinfo{author}{Hirose, T.},
  \bibinfo{author}{Ikari, M.J.}, \bibinfo{author}{Ishikawa, T.},
  \bibinfo{author}{Jeppson, T.}, \bibinfo{author}{Kano, Y.},
  \bibinfo{author}{Kirkpatrick, J.}, \bibinfo{author}{Kodaira, S.},
  \bibinfo{author}{Lin, W.}, \bibinfo{author}{Nakamura, Y.},
  \bibinfo{author}{Rabinowitz, H.S.}, \bibinfo{author}{Regalla, C.},
  \bibinfo{author}{Remitti, F.}, \bibinfo{author}{Rowe, C.},
  \bibinfo{author}{Saffer, D.M.}, \bibinfo{author}{Saito, S.},
  \bibinfo{author}{Sample, J.}, \bibinfo{author}{Sanada, Y.},
  \bibinfo{author}{Savage, H.M.}, \bibinfo{author}{Sun, T.},
  \bibinfo{author}{Toczko, S.}, \bibinfo{author}{Ujiie, K.},
  \bibinfo{author}{Wolfson-Schwehr, M.}, \bibinfo{author}{Yang, T.},
  \bibinfo{year}{2020}.
\newblock \bibinfo{title}{The state of stress on the fault before, during, and
  after a major earthquake}.
\newblock \bibinfo{journal}{Annual Review of Earth and Planetary Sciences}
  \bibinfo{volume}{48}, \bibinfo{pages}{49--74}.
\newblock \DOIprefix\doi{10.1146/annurev-earth-053018-060507}.
\bibitem[{Byrd et~al.(1995)Byrd, Lu, Nocedal and Zhu}]{byrd1995limited}
\bibinfo{author}{Byrd, R.H.}, \bibinfo{author}{Lu, P.},
  \bibinfo{author}{Nocedal, J.}, \bibinfo{author}{Zhu, C.},
  \bibinfo{year}{1995}.
\newblock \bibinfo{title}{A limited memory algorithm for bound constrained
  optimization}.
\newblock \bibinfo{journal}{SIAM Journal on scientific computing}
  \bibinfo{volume}{16}, \bibinfo{pages}{1190--1208}.
\newblock \DOIprefix\doi{10.1137/0916069}.
\bibitem[{Cai et~al.(2021)Cai, Wang, Wang, Perdikaris and
  Karniadakis}]{cai2021physics}
\bibinfo{author}{Cai, S.}, \bibinfo{author}{Wang, Z.}, \bibinfo{author}{Wang,
  S.}, \bibinfo{author}{Perdikaris, P.}, \bibinfo{author}{Karniadakis, G.E.},
  \bibinfo{year}{2021}.
\newblock \bibinfo{title}{Physics-informed neural networks for heat transfer
  problems}.
\newblock \bibinfo{journal}{Journal of Heat Transfer} \bibinfo{volume}{143}.
\newblock \DOIprefix\doi{10.1115/1.4050542}.
\bibitem[{Ciarlet(2002)}]{ciarlet2002finite}
\bibinfo{author}{Ciarlet, P.G.}, \bibinfo{year}{2002}.
\newblock \bibinfo{title}{The finite element method for elliptic problems}.
  volume~\bibinfo{volume}{40}.
\newblock \bibinfo{publisher}{Siam}.
\bibitem[{Cuomo et~al.(2022)Cuomo, Di~Cola, Giampaolo, Rozza, Raissi and
  Piccialli}]{cuomo2022scientific}
\bibinfo{author}{Cuomo, S.}, \bibinfo{author}{Di~Cola, V.S.},
  \bibinfo{author}{Giampaolo, F.}, \bibinfo{author}{Rozza, G.},
  \bibinfo{author}{Raissi, M.}, \bibinfo{author}{Piccialli, F.},
  \bibinfo{year}{2022}.
\newblock \bibinfo{title}{Scientific machine learning through physics--informed
  neural networks: Where we are and what’s next}.
\newblock \bibinfo{journal}{Journal of Scientific Computing}
  \bibinfo{volume}{92}, \bibinfo{pages}{88}.
\newblock \DOIprefix\doi{10.1007/s10915-022-01939-z}.
\bibitem[{De~Ryck et~al.(2022)De~Ryck, Jagtap and Mishra}]{de2022error}
\bibinfo{author}{De~Ryck, T.}, \bibinfo{author}{Jagtap, A.D.},
  \bibinfo{author}{Mishra, S.}, \bibinfo{year}{2022}.
\newblock \bibinfo{title}{Error estimates for physics informed neural networks
  approximating the navier-stokes equations}.
\newblock \bibinfo{journal}{arXiv preprint arXiv:2203.09346}
  \DOIprefix\doi{10.48550/arXiv.2203.09346}.
\bibitem[{De~Ryck et~al.(2021)De~Ryck, Lanthaler and
  Mishra}]{de2021approximation}
\bibinfo{author}{De~Ryck, T.}, \bibinfo{author}{Lanthaler, S.},
  \bibinfo{author}{Mishra, S.}, \bibinfo{year}{2021}.
\newblock \bibinfo{title}{On the approximation of functions by tanh neural
  networks}.
\newblock \bibinfo{journal}{Neural Networks} \bibinfo{volume}{143},
  \bibinfo{pages}{732--750}.
\newblock \DOIprefix\doi{10.1016/j.neunet.2021.08.015}.
\bibitem[{Dieterich(1979)}]{dieterich1979}
\bibinfo{author}{Dieterich, J.H.}, \bibinfo{year}{1979}.
\newblock \bibinfo{title}{Modeling of rock friction 1. {{Experimental}} results
  and constitutive equations}.
\newblock \bibinfo{journal}{Journal of Geophysical Research, [Solid Earth]}
  \bibinfo{volume}{84}, \bibinfo{pages}{2161--2168}.
\newblock \DOIprefix\doi{10.1029/JB084iB05p02161}.
\bibitem[{Erickson and Dunham(2014)}]{Erickson2014}
\bibinfo{author}{Erickson, B.A.}, \bibinfo{author}{Dunham, E.M.},
  \bibinfo{year}{2014}.
\newblock \bibinfo{title}{An efficient numerical method for earthquake cycles
  in heterogeneous media: Alternating subbasin and surface-rupturing events on
  faults crossing a sedimentary basin}.
\newblock \bibinfo{journal}{Journal of Geophysical Research-Solid Earth}
  \bibinfo{volume}{119}, \bibinfo{pages}{3290--3316}.
\newblock \DOIprefix\doi{10.1002/2013JB010614}.
\bibitem[{Erickson et~al.(2020)Erickson, Jiang, Barall, Lapusta, Dunham,
  Harris, Abrahams, Allison, Ampuero, Barbot, Cattania, Elbanna, Fialko, Idini,
  Kozdon, Lambert, Liu, Luo, Ma, Mckay, Segall, Shi, van~den Ende and
  Wei}]{Erickson2020}
\bibinfo{author}{Erickson, B.A.}, \bibinfo{author}{Jiang, J.},
  \bibinfo{author}{Barall, M.}, \bibinfo{author}{Lapusta, N.},
  \bibinfo{author}{Dunham, E.M.}, \bibinfo{author}{Harris, R.},
  \bibinfo{author}{Abrahams, L.S.}, \bibinfo{author}{Allison, K.L.},
  \bibinfo{author}{Ampuero, J.P.}, \bibinfo{author}{Barbot, S.},
  \bibinfo{author}{Cattania, C.}, \bibinfo{author}{Elbanna, A.},
  \bibinfo{author}{Fialko, Y.}, \bibinfo{author}{Idini, B.},
  \bibinfo{author}{Kozdon, J.E.}, \bibinfo{author}{Lambert, V.},
  \bibinfo{author}{Liu, Y.}, \bibinfo{author}{Luo, Y.}, \bibinfo{author}{Ma,
  X.}, \bibinfo{author}{Mckay, M.B.}, \bibinfo{author}{Segall, P.},
  \bibinfo{author}{Shi, P.}, \bibinfo{author}{van~den Ende, M.},
  \bibinfo{author}{Wei, M.}, \bibinfo{year}{2020}.
\newblock \bibinfo{title}{The community code verification exercise for
  simulating sequences of earthquakes and aseismic slip ({SEAS})}.
\newblock \bibinfo{journal}{Seismol. Res. Lett.} \bibinfo{volume}{91},
  \bibinfo{pages}{874--890}.
\newblock \DOIprefix\doi{10.1785/0220190248}.
\bibitem[{{Erickson Laboratory}()}]{EQP}
\bibinfo{author}{{Erickson Laboratory}}, .
\newblock \bibinfo{title}{Earthquake {PINN}s}.
\newblock \bibinfo{howpublished}{{https://github.com/Thrase/EQ\_pinns}}.
\newblock \bibinfo{note}{Accessed: 2023-09-16}.
\bibitem[{Ern and Guermond(2004)}]{ern2004theory}
\bibinfo{author}{Ern, A.}, \bibinfo{author}{Guermond, J.L.},
  \bibinfo{year}{2004}.
\newblock \bibinfo{title}{Theory and practice of finite elements}. volume
  \bibinfo{volume}{159}.
\newblock \bibinfo{publisher}{Springer}.
\newblock \DOIprefix\doi{10.1007/978-1-4757-4355-5}.
\bibitem[{Fukushima et~al.(2023)Fukushima, Kano and Hirahara}]{Fukushima2023}
\bibinfo{author}{Fukushima, R.}, \bibinfo{author}{Kano, M.},
  \bibinfo{author}{Hirahara, K.}, \bibinfo{year}{2023}.
\newblock \bibinfo{title}{Physics-informed neural networks for fault slip
  monitoring: simulation, frictional parameter estimation, and prediction on
  slow slip events in a spring-slider system}.
\newblock \bibinfo{journal}{ESS Open Archive}
  \DOIprefix\doi{10.22541/essoar.168988460.01601423/v1}.
\bibitem[{Givoli(2021)}]{givoli2021tutorial}
\bibinfo{author}{Givoli, D.}, \bibinfo{year}{2021}.
\newblock \bibinfo{title}{A tutorial on the adjoint method for inverse
  problems}.
\newblock \bibinfo{journal}{Computer Methods in Applied Mechanics and
  Engineering} \bibinfo{volume}{380}, \bibinfo{pages}{113810}.
\newblock \DOIprefix\doi{10.1016/j.cma.2021.113810}.
\bibitem[{Glorot and Bengio(2010)}]{glorot2010understanding}
\bibinfo{author}{Glorot, X.}, \bibinfo{author}{Bengio, Y.},
  \bibinfo{year}{2010}.
\newblock \bibinfo{title}{Understanding the difficulty of training deep
  feedforward neural networks}, in: \bibinfo{booktitle}{Proceedings of the
  thirteenth international conference on artificial intelligence and
  statistics}, \bibinfo{organization}{JMLR Workshop and Conference
  Proceedings}. pp. \bibinfo{pages}{249--256}.
\bibitem[{Goswami et~al.(2020)Goswami, Anitescu, Chakraborty and
  Rabczuk}]{goswami2020transfer}
\bibinfo{author}{Goswami, S.}, \bibinfo{author}{Anitescu, C.},
  \bibinfo{author}{Chakraborty, S.}, \bibinfo{author}{Rabczuk, T.},
  \bibinfo{year}{2020}.
\newblock \bibinfo{title}{Transfer learning enhanced physics informed neural
  network for phase-field modeling of fracture}.
\newblock \bibinfo{journal}{Theoretical and Applied Fracture Mechanics}
  \bibinfo{volume}{106}, \bibinfo{pages}{102447}.
\newblock \DOIprefix\doi{10.1016/j.tafmec.2019.102447}.
\bibitem[{Haghighat et~al.(2021)Haghighat, Raissi, Moure, Gomez and
  Juanes}]{haghighat2021physics}
\bibinfo{author}{Haghighat, E.}, \bibinfo{author}{Raissi, M.},
  \bibinfo{author}{Moure, A.}, \bibinfo{author}{Gomez, H.},
  \bibinfo{author}{Juanes, R.}, \bibinfo{year}{2021}.
\newblock \bibinfo{title}{A physics-informed deep learning framework for
  inversion and surrogate modeling in solid mechanics}.
\newblock \bibinfo{journal}{Computer Methods in Applied Mechanics and
  Engineering} \bibinfo{volume}{379}, \bibinfo{pages}{113741}.
\newblock \DOIprefix\doi{10.1016/j.cma.2021.113741}.
\bibitem[{Harris et~al.(2009)Harris, Barall, Archuleta, Dunham, Aagaard,
  Ampuero, Bhat, Cruz-Atienza, Dalguer, Dawson, Day, Duan, Ely, Kaneko, Kase,
  Lapusta, Liu, Ma, Oglesby, Olsen, Pitarka, Song and Templeton}]{Harris2009}
\bibinfo{author}{Harris, R.A.}, \bibinfo{author}{Barall, M.},
  \bibinfo{author}{Archuleta, R.}, \bibinfo{author}{Dunham, E.M.},
  \bibinfo{author}{Aagaard, B.}, \bibinfo{author}{Ampuero, J.P.},
  \bibinfo{author}{Bhat, H.}, \bibinfo{author}{Cruz-Atienza, V.},
  \bibinfo{author}{Dalguer, L.}, \bibinfo{author}{Dawson, P.},
  \bibinfo{author}{Day, S.}, \bibinfo{author}{Duan, B.}, \bibinfo{author}{Ely,
  G.}, \bibinfo{author}{Kaneko, Y.}, \bibinfo{author}{Kase, Y.},
  \bibinfo{author}{Lapusta, N.}, \bibinfo{author}{Liu, Y.},
  \bibinfo{author}{Ma, S.}, \bibinfo{author}{Oglesby, D.},
  \bibinfo{author}{Olsen, K.}, \bibinfo{author}{Pitarka, A.},
  \bibinfo{author}{Song, S.}, \bibinfo{author}{Templeton, E.},
  \bibinfo{year}{2009}.
\newblock \bibinfo{title}{The {SCEC/USGS} dynamic earthquake rupture code
  verification exercise}.
\newblock \bibinfo{journal}{Seismol. Res. Lett.} \bibinfo{volume}{80},
  \bibinfo{pages}{119--126}.
\newblock \DOIprefix\doi{10.1785/gssrl.80.1.119}.
\bibitem[{Harvey et~al.(2023)Harvey, Erickson and Kozdon}]{Harvey2022}
\bibinfo{author}{Harvey, T.W.}, \bibinfo{author}{Erickson, B.A.},
  \bibinfo{author}{Kozdon, J.E.}, \bibinfo{year}{2023}.
\newblock \bibinfo{title}{A high-order accurate summation-by-parts finite
  difference method for fully-dynamic earthquake sequence simulations within
  sedimentary basins}.
\newblock \bibinfo{journal}{Journal of Geophysical Research: Solid Earth} ,
  \bibinfo{pages}{e2022JB025357}\DOIprefix\doi{10.1029/2022JB025357}.
\bibitem[{Ide et~al.(2007)Ide, Beroza, Shelly and Uchide}]{Ide2007}
\bibinfo{author}{Ide, S.}, \bibinfo{author}{Beroza, G.C.},
  \bibinfo{author}{Shelly, D.R.}, \bibinfo{author}{Uchide, T.},
  \bibinfo{year}{2007}.
\newblock \bibinfo{title}{A scaling law for slow earthquakes}.
\newblock \bibinfo{journal}{Nature} \bibinfo{volume}{447},
  \bibinfo{pages}{76--79}.
\bibitem[{Jagtap et~al.(2020)Jagtap, Kharazmi and
  Karniadakis}]{jagtap2020conservative}
\bibinfo{author}{Jagtap, A.D.}, \bibinfo{author}{Kharazmi, E.},
  \bibinfo{author}{Karniadakis, G.E.}, \bibinfo{year}{2020}.
\newblock \bibinfo{title}{Conservative physics-informed neural networks on
  discrete domains for conservation laws: Applications to forward and inverse
  problems}.
\newblock \bibinfo{journal}{Computer Methods in Applied Mechanics and
  Engineering} \bibinfo{volume}{365}, \bibinfo{pages}{113028}.
\newblock \DOIprefix\doi{10.1016/j.cma.2020.113028}.
\bibitem[{Jin et~al.(2021)Jin, Cai, Li and Karniadakis}]{jin2021nsfnets}
\bibinfo{author}{Jin, X.}, \bibinfo{author}{Cai, S.}, \bibinfo{author}{Li, H.},
  \bibinfo{author}{Karniadakis, G.E.}, \bibinfo{year}{2021}.
\newblock \bibinfo{title}{Nsfnets (navier-stokes flow nets): Physics-informed
  neural networks for the incompressible navier-stokes equations}.
\newblock \bibinfo{journal}{Journal of Computational Physics}
  \bibinfo{volume}{426}, \bibinfo{pages}{109951}.
\newblock \DOIprefix\doi{https://doi.org/10.1016/j.jcp.2020.109951}.
\bibitem[{Karimpouli and Tahmasebi(2020)}]{KARIMPOULI20201993}
\bibinfo{author}{Karimpouli, S.}, \bibinfo{author}{Tahmasebi, P.},
  \bibinfo{year}{2020}.
\newblock \bibinfo{title}{Physics informed machine learning: Seismic wave
  equation}.
\newblock \bibinfo{journal}{Geoscience Frontiers} \bibinfo{volume}{11},
  \bibinfo{pages}{1993--2001}.
\newblock \DOIprefix\doi{10.1016/j.gsf.2020.07.007}.
\bibitem[{Karniadakis et~al.(2021)Karniadakis, Kevrekidis, Lu, Perdikaris, Wang
  and Yang}]{karniadakis2021physics}
\bibinfo{author}{Karniadakis, G.E.}, \bibinfo{author}{Kevrekidis, I.G.},
  \bibinfo{author}{Lu, L.}, \bibinfo{author}{Perdikaris, P.},
  \bibinfo{author}{Wang, S.}, \bibinfo{author}{Yang, L.}, \bibinfo{year}{2021}.
\newblock \bibinfo{title}{Physics-informed machine learning}.
\newblock \bibinfo{journal}{Nature Reviews Physics} \bibinfo{volume}{3},
  \bibinfo{pages}{422--440}.
\newblock \DOIprefix\doi{10.1038/s42254-021-00314-5}.
\bibitem[{Kern(2016)}]{kern2016numerical}
\bibinfo{author}{Kern, M.}, \bibinfo{year}{2016}.
\newblock \bibinfo{title}{Numerical methods for inverse problems}.
\newblock \bibinfo{publisher}{John Wiley \& Sons}.
\bibitem[{Kharazmi et~al.(2019)Kharazmi, Zhang and
  Karniadakis}]{kharazmi2019variational}
\bibinfo{author}{Kharazmi, E.}, \bibinfo{author}{Zhang, Z.},
  \bibinfo{author}{Karniadakis, G.E.}, \bibinfo{year}{2019}.
\newblock \bibinfo{title}{Variational physics-informed neural networks for
  solving partial differential equations}.
\newblock \bibinfo{journal}{arXiv preprint arXiv:1912.00873}
  \DOIprefix\doi{10.48550/arXiv.1912.00873}.
\bibitem[{Kharazmi et~al.(2021)Kharazmi, Zhang and
  Karniadakis}]{kharazmi2021hp}
\bibinfo{author}{Kharazmi, E.}, \bibinfo{author}{Zhang, Z.},
  \bibinfo{author}{Karniadakis, G.E.}, \bibinfo{year}{2021}.
\newblock \bibinfo{title}{hp-vpinns: Variational physics-informed neural
  networks with domain decomposition}.
\newblock \bibinfo{journal}{Computer Methods in Applied Mechanics and
  Engineering} \bibinfo{volume}{374}, \bibinfo{pages}{113547}.
\newblock \DOIprefix\doi{10.1016/j.cma.2020.113547}.
\bibitem[{Kollmannsberger et~al.(2021)Kollmannsberger, D’Angella, Jokeit,
  Herrmann et~al.}]{kollmannsberger2021deep}
\bibinfo{author}{Kollmannsberger, S.}, \bibinfo{author}{D’Angella, D.},
  \bibinfo{author}{Jokeit, M.}, \bibinfo{author}{Herrmann, L.}, et~al.,
  \bibinfo{year}{2021}.
\newblock \bibinfo{title}{Deep Learning in Computational Mechanics}.
\newblock \bibinfo{publisher}{Springer}.
\bibitem[{Kong et~al.(2018)Kong, Trugman, Ross, Bianco, Meade and
  Gerstoft}]{Kong2018}
\bibinfo{author}{Kong, Q.}, \bibinfo{author}{Trugman, D.T.},
  \bibinfo{author}{Ross, Z.E.}, \bibinfo{author}{Bianco, M.J.},
  \bibinfo{author}{Meade, B.J.}, \bibinfo{author}{Gerstoft, P.},
  \bibinfo{year}{2018}.
\newblock \bibinfo{title}{{Machine Learning in Seismology: Turning Data into
  Insights}}.
\newblock \bibinfo{journal}{Seismological Research Letters}
  \bibinfo{volume}{90}, \bibinfo{pages}{3--14}.
\newblock \DOIprefix\doi{10.1785/0220180259}.
\bibitem[{Kutyniok(2022)}]{kutyniok2022mathematics}
\bibinfo{author}{Kutyniok, G.}, \bibinfo{year}{2022}.
\newblock \bibinfo{title}{The mathematics of artificial intelligence}.
\newblock \bibinfo{journal}{arXiv preprint arXiv:2203.08890}
  \DOIprefix\doi{10.48550/arXiv.2203.08890}.
\bibitem[{Lagaris et~al.(1998)Lagaris, Likas and
  Fotiadis}]{lagaris1998artificial}
\bibinfo{author}{Lagaris, I.E.}, \bibinfo{author}{Likas, A.},
  \bibinfo{author}{Fotiadis, D.I.}, \bibinfo{year}{1998}.
\newblock \bibinfo{title}{Artificial neural networks for solving ordinary and
  partial differential equations}.
\newblock \bibinfo{journal}{IEEE transactions on neural networks}
  \bibinfo{volume}{9}, \bibinfo{pages}{987--1000}.
\newblock \DOIprefix\doi{10.1109/72.712178}.
\bibitem[{Lagaris et~al.(2000)Lagaris, Likas and
  Papageorgiou}]{lagaris2000neural}
\bibinfo{author}{Lagaris, I.E.}, \bibinfo{author}{Likas, A.C.},
  \bibinfo{author}{Papageorgiou, D.G.}, \bibinfo{year}{2000}.
\newblock \bibinfo{title}{Neural-network methods for boundary value problems
  with irregular boundaries}.
\newblock \bibinfo{journal}{IEEE Transactions on Neural Networks}
  \bibinfo{volume}{11}, \bibinfo{pages}{1041--1049}.
\newblock \DOIprefix\doi{10.1109/72.870037}.
\bibitem[{Mao et~al.(2020)Mao, Jagtap and Karniadakis}]{mao2020physics}
\bibinfo{author}{Mao, Z.}, \bibinfo{author}{Jagtap, A.D.},
  \bibinfo{author}{Karniadakis, G.E.}, \bibinfo{year}{2020}.
\newblock \bibinfo{title}{Physics-informed neural networks for high-speed
  flows}.
\newblock \bibinfo{journal}{Computer Methods in Applied Mechanics and
  Engineering} \bibinfo{volume}{360}, \bibinfo{pages}{112789}.
\newblock \DOIprefix\doi{10.1016/j.cma.2019.112789}.
\bibitem[{Marone(1998)}]{Marone1998}
\bibinfo{author}{Marone, C.}, \bibinfo{year}{1998}.
\newblock \bibinfo{title}{Laboratory-derived friction laws and their
  application to seismic faulting}.
\newblock \bibinfo{journal}{Ann. Rev. Earth Pl. Sc.} \bibinfo{volume}{26},
  \bibinfo{pages}{643--696}.
\newblock \DOIprefix\doi{10.1146/annurev.earth.26.1.643}.
\bibitem[{Marone and Kilgore(1993)}]{Marone1993}
\bibinfo{author}{Marone, C.}, \bibinfo{author}{Kilgore, B.},
  \bibinfo{year}{1993}.
\newblock \bibinfo{title}{Scaling of the critical slip distance for seismic
  faulting with shear strain in fault zones}.
\newblock \bibinfo{journal}{Nature} \bibinfo{volume}{362},
  \bibinfo{pages}{618--621}.
\newblock \DOIprefix\doi{10.1038/362618a0}.
\bibitem[{Mishra and Molinaro(2022a)}]{mishra2022estimates_inverse}
\bibinfo{author}{Mishra, S.}, \bibinfo{author}{Molinaro, R.},
  \bibinfo{year}{2022}a.
\newblock \bibinfo{title}{Estimates on the generalization error of
  physics-informed neural networks for approximating a class of inverse
  problems for pdes}.
\newblock \bibinfo{journal}{IMA Journal of Numerical Analysis}
  \bibinfo{volume}{42}, \bibinfo{pages}{981--1022}.
\newblock \DOIprefix\doi{10.1093/imanum/drab032}.
\bibitem[{Mishra and Molinaro(2022b)}]{mishra2022estimates_forward}
\bibinfo{author}{Mishra, S.}, \bibinfo{author}{Molinaro, R.},
  \bibinfo{year}{2022}b.
\newblock \bibinfo{title}{Estimates on the generalization error of
  physics-informed neural networks for approximating pdes}.
\newblock \bibinfo{journal}{IMA Journal of Numerical Analysis}
  \DOIprefix\doi{10.1093/imanum/drab093}.
\bibitem[{M{\"u}ller and Zeinhofer(2023)}]{muller2023achieving}
\bibinfo{author}{M{\"u}ller, J.}, \bibinfo{author}{Zeinhofer, M.},
  \bibinfo{year}{2023}.
\newblock \bibinfo{title}{Achieving high accuracy with pinns via energy natural
  gradients}.
\newblock \bibinfo{journal}{arXiv preprint arXiv:2302.13163}
  \DOIprefix\doi{10.48550/arXiv.2302.13163}.
\bibitem[{{National Academies of Sciences, Engineering, and Medicine and
  others}(2020)}]{NAP25761}
\bibinfo{author}{{National Academies of Sciences, Engineering, and Medicine and
  others}}, \bibinfo{year}{2020}.
\newblock \bibinfo{title}{A vision for NSF Earth sciences 2020-2030: Earth in
  time}.
\newblock \bibinfo{publisher}{National Academies Press},
  \bibinfo{address}{Washington, DC}.
\bibitem[{Okazaki et~al.(2022)Okazaki, Ito, Hirahara and Ueda}]{Okazaki2022}
\bibinfo{author}{Okazaki, T.}, \bibinfo{author}{Ito, T.},
  \bibinfo{author}{Hirahara, K.}, \bibinfo{author}{Ueda, N.},
  \bibinfo{year}{2022}.
\newblock \bibinfo{title}{Physics-informed deep learning approach for modeling
  crustal deformation}.
\newblock \bibinfo{journal}{Nature Communications} \bibinfo{volume}{13},
  \bibinfo{pages}{7092}.
\newblock \DOIprefix\doi{10.1038/s41467-022-34922-1}.
\bibitem[{Raissi et~al.(2019)Raissi, Perdikaris and
  Karniadakis}]{raissi2019physics}
\bibinfo{author}{Raissi, M.}, \bibinfo{author}{Perdikaris, P.},
  \bibinfo{author}{Karniadakis, G.E.}, \bibinfo{year}{2019}.
\newblock \bibinfo{title}{Physics-informed neural networks: A deep learning
  framework for solving forward and inverse problems involving nonlinear
  partial differential equations}.
\newblock \bibinfo{journal}{Journal of Computational physics}
  \bibinfo{volume}{378}, \bibinfo{pages}{686--707}.
\newblock \DOIprefix\doi{10.1016/j.jcp.2018.10.045}.
\bibitem[{Roache(1998)}]{Roache}
\bibinfo{author}{Roache, P.}, \bibinfo{year}{1998}.
\newblock \bibinfo{title}{Verification and Validation in Computational Science
  and Engineering}.
\newblock \bibinfo{edition}{1} ed., \bibinfo{publisher}{Hermosa Publishers},
  \bibinfo{address}{Albuquerque, NM}.
\bibitem[{Ruina(1983)}]{Ruina1983}
\bibinfo{author}{Ruina, A.}, \bibinfo{year}{1983}.
\newblock \bibinfo{title}{Slip instability and state variable friction laws}.
\newblock \bibinfo{journal}{J. Geophys. Res. Solid Earth} \bibinfo{volume}{88},
  \bibinfo{pages}{10359--10370}.
\newblock \DOIprefix\doi{10.1029/JB088iB12p10359}.
\bibitem[{Scholz(2019)}]{scholz_2019}
\bibinfo{author}{Scholz, C.H.}, \bibinfo{year}{2019}.
\newblock \bibinfo{title}{The Mechanics of Earthquakes and Faulting}.
\newblock \bibinfo{edition}{3} ed., \bibinfo{publisher}{Cambridge University
  Press}.
\newblock \DOIprefix\doi{10.1017/9781316681473}.
\bibitem[{Shin and Em~Karniadakis(2020)}]{shin2020convergence}
\bibinfo{author}{Shin, YeonjongDarbon, J.}, \bibinfo{author}{Em~Karniadakis,
  G.}, \bibinfo{year}{2020}.
\newblock \bibinfo{title}{On the convergence of physics informed neural
  networks for linear second-order elliptic and parabolic type pdes}.
\newblock \bibinfo{journal}{Communications in Computational Physics}
  \bibinfo{volume}{28}, \bibinfo{pages}{2042--2074}.
\newblock \DOIprefix\doi{10.4208/cicp.OA-2020-0193}.
\bibitem[{Shin et~al.(2023)Shin, Zhang and Karniadakis}]{shin2020error}
\bibinfo{author}{Shin, Y.}, \bibinfo{author}{Zhang, Z.},
  \bibinfo{author}{Karniadakis, G.E.}, \bibinfo{year}{2023}.
\newblock \bibinfo{title}{Error estimates of residual minimization using neural
  networks for linear pdes}.
\newblock \bibinfo{journal}{Journal of Machine Learning for Modeling and
  Computing} \bibinfo{volume}{4}, \bibinfo{pages}{73--101}.
\newblock \DOIprefix\doi{10.1615/JMachLearnModelComput.2023050411}.
\bibitem[{Sun et~al.(2020)Sun, Gao, Pan and Wang}]{sun2020surrogate}
\bibinfo{author}{Sun, L.}, \bibinfo{author}{Gao, H.}, \bibinfo{author}{Pan,
  S.}, \bibinfo{author}{Wang, J.X.}, \bibinfo{year}{2020}.
\newblock \bibinfo{title}{Surrogate modeling for fluid flows based on
  physics-constrained deep learning without simulation data}.
\newblock \bibinfo{journal}{Computer Methods in Applied Mechanics and
  Engineering} \bibinfo{volume}{361}, \bibinfo{pages}{112732}.
\newblock \DOIprefix\doi{10.1016/j.cma.2019.112732}.
\bibitem[{{van den Ende} et~al.(2018){van den Ende}, Chen, Ampuero and
  Niemeijer}]{VANDENENDE2018273}
\bibinfo{author}{{van den Ende}, M.}, \bibinfo{author}{Chen, J.},
  \bibinfo{author}{Ampuero, J.P.}, \bibinfo{author}{Niemeijer, A.},
  \bibinfo{year}{2018}.
\newblock \bibinfo{title}{A comparison between rate-and-state friction and
  microphysical models, based on numerical simulations of fault slip}.
\newblock \bibinfo{journal}{Tectonophysics} \bibinfo{volume}{733},
  \bibinfo{pages}{273--295}.
\newblock \DOIprefix\doi{10.1016/j.tecto.2017.11.040}.
\bibitem[{Wang and Cai(2020)}]{wang2020multi}
\bibinfo{author}{Wang, BoZhang, W.}, \bibinfo{author}{Cai, W.},
  \bibinfo{year}{2020}.
\newblock \bibinfo{title}{Multi-scale deep neural network (mscalednn) methods
  for oscillatory stokes flows in complex domains}.
\newblock \bibinfo{journal}{Communications in Computational Physics}
  \bibinfo{volume}{28}, \bibinfo{pages}{2139--2157}.
\newblock \DOIprefix\doi{10.4208/cicp.OA-2020-0192}.
\bibitem[{Yu et~al.(2018)}]{yu2018deep}
\bibinfo{author}{Yu, B.}, et~al., \bibinfo{year}{2018}.
\newblock \bibinfo{title}{The deep ritz method: a deep learning-based numerical
  algorithm for solving variational problems}.
\newblock \bibinfo{journal}{Communications in Mathematics and Statistics}
  \bibinfo{volume}{6}, \bibinfo{pages}{1--12}.
\newblock \DOIprefix\doi{10.1007/s40304-018-0127-z}.
\bibitem[{Zhao et~al.(2023)Zhao, Ling, Liu, Wang, Burke and Lian}]{Zhao2023}
\bibinfo{author}{Zhao, J.}, \bibinfo{author}{Ling, H.}, \bibinfo{author}{Liu,
  J.}, \bibinfo{author}{Wang, J.}, \bibinfo{author}{Burke, A.F.},
  \bibinfo{author}{Lian, Y.}, \bibinfo{year}{2023}.
\newblock \bibinfo{title}{Machine learning for predicting battery capacity for
  electric vehicles}.
\newblock \bibinfo{journal}{eTransportation} \bibinfo{volume}{15},
  \bibinfo{pages}{100214}.
\newblock \DOIprefix\doi{10.1016/j.etran.2022.100214}.

\end{thebibliography}

\end{document}